\documentclass[5p]{elsarticle}

\usepackage{natbib}
\setcitestyle{authoryear,open={(},close={)}} %Citations bracketed with ()
\setcitestyle{citesep={;}} %Makes multiple citations with ;
\usepackage{graphicx}
\usepackage{footnote}
\makesavenoteenv{tabular}
\makesavenoteenv{table}
\usepackage{multicol}
\usepackage{cprotect}

\usepackage[printwatermark]{xwatermark}
\usepackage{xcolor, colortbl}
\usepackage{tikz}
\usepackage{url}
\usepackage{amssymb}
\usepackage{dcolumn}% Align table columns on decimal point
\usepackage{bm}% bold math
\usepackage[switch]{lineno}
\usepackage{caption}
\usepackage{subcaption}
\usepackage{amsmath}

\DeclareCaptionType{equ}[Equation][List of Equations]
\journal{Astronomy And Computing}

\begin{document}\sloppy
\definecolor{Gray}{gray}{0.9}
\begin{frontmatter}


\title{Scalability Model for the LOFAR Direction Independent Pipeline}% Force line breaks with \\
%speedup real life pipelines performance automatic measurements radio astronomy
%\thanks{A footnote to the article title}%X
\author{A.P. Mechev $^a$}
\ead{apmechev@strw.leidenuniv.nl}

\author{T.W. Shimwell $^b$}%
\author{A. Plaat $^c$}%
\author{H. Intema $^{ad}$}%
\author{A.L. Varbanescu $^e$}
\author{H.J.A Rottgering $^a$}%

\date{\today} 
\address{$^a$ Leiden Observatory, Niels Bohrweg 2, 2333 CA Leiden, the Netherlands}
\address{$^b$ ASTRON, Oude Hoogeveensedijk 4, 7991 PD , The Netherlands }
\address{$^c$ Leiden Institute of Advanced Computer Science, Niels Bohrweg 1, 2333 CA Leiden, the Netherlands}
\address{$^d$ International Centre for Radio Astronomy Research -- Curtin University, GPO Box U1987, Perth, WA 6845, Australia}
\address{$^e$ University of Amsterdam, Spui 21, 1012 WX Amsterdam, the Netherlands}

\begin{abstract}
LOFAR is a leading aperture synthesis telescope operated in the Netherlands with stations across Europe. The LOFAR Two-meter Sky Survey (LoTSS) will produce more than 3000 14 TB data sets, mapping the entire northern sky at low frequencies. The data produced by this survey is important for understanding the formation and evolution of galaxies, supermassive black holes and other astronomical phenomena. All of the LoTSS data needs to be processed by the LOFAR Direction Independent (DI) pipeline, \texttt{prefactor}. Understanding the performance of this pipeline is important when trying to optimize the throughput for  large projects, such as LoTSS and other deep surveys. Making a model of its completion time will enable us to predict the time taken to process large data sets, optimize our parameter choices, help schedule other LOFAR processing services, and predict processing time for future large radio telescopes. We tested the \texttt{prefactor} pipeline by scaling several parameters, notably number of CPUs, data size and size of calibration sky model. We present these results as a comprehensive model which will be used to predict processing time for a wide range of processing parameters. We also discover that smaller calibration models lead to significantly faster calibration times, while the calibration results do not significantly degrade in quality. Finally, we validate the model and compare predictions with production runs from the past six months, quantifying the performance penalties incurred by processing on a shared cluster. We conclude by noting the utility of the results and model for the LoTSS Survey, LOFAR as a whole and for other telescopes. 

% \item[Prefactor]
% The \texttt{LOFAR} pre-factor pipeline prepares LOFAR Observations for creating high fidelity images. 
% \end{description}
\end{abstract}
\end{frontmatter}

%\maketitle
%
%\tableofcontents
%\linenumbers

\section{\label{sec:intro}Introduction }

Astronomy has entered the big data era with many projects creating petabytes of data per year. This data is often processed by complex multi-step pipelines consisting of various algorithms. Understanding the scalability of astronomical algorithms theoretically, in a controlled environment, and in production is important for making predictions for the data reduction of future projects and upcoming telescopes. 

The Low Frequency Array (LOFAR) \citep{LOFAR} is a leading European low-frequency radio telescope. The majority of LOFAR's stations are in the Netherlands, however the telescope can use stations across Europe to create ultra-high resolution radio maps. LOFAR data needs to undergo several computationally intensive processing steps before obtaining a final scientific image. 

To create a broadband image, LOFAR data is first processed by a Direction Independent (DI) Calibration pipeline followed by Direction Dependent (DD) Calibration pipeline \citep[e.g.][]{lofar_prefactor, Wendy_bootes,tassesmirnov, tasse2018faceting}. The goal of DI calibration is to remove effects that are constant across the target field such as radio frequency interference, contamination by bright off-axis sources and antenna gains. After this step, DD Calibration focuses on removing effects which vary across the field, such as ionospheric and beam effects. The result of these two pipelines is a science-ready image. 

Our implementation of the DI LOFAR processing, \texttt{prefactor}, can be parallelized on a high throughput cluster \citep{mechev17}. The Direction Dependent processing, implemented in \texttt{ddf-pipeline}\footnote{Available at \raggedright\href{https://github.com/mhardcastle/ddf-pipeline/releases}{https://github.com/mhardcastle/ddf-pipeline/releases}}, is subsequently performed on a single HPC node. 

The LOFAR Surveys Key Science Project (SKSP) \citep{lotss, LOTSS_DR2} is a long running project consisting of several low frequency surveys of the northern sky. The broadest tier of the survey, LoTSS, will use more than 3000 8-hour observations to create maps with a noise levels below 100 $\mu$Jy. We have already processed more than 500 of these observations using the \texttt{prefactor} DI pipeline \citep{lofar_prefactor, prefactor_zenodo}. 

While the current LoTSS imaging algorithms can process data averaged by up to a factor of 64 in frequency and time, it is important to understand how LOFAR processing scales with processing parameters, such as averaging parameters. Since LOFAR data is used by multiple scientific teams, not every team can produce scientific results from data averaged by such a high factor. Users from those teams need to be able to predict the time and computational resources required to process their data, taking into account the increasing LOFAR observation rates, data sizes and scientific requirements. 

We study the scalability of processing LOFAR data, by setting up processing of a sample SKSP data set on an isolated node on the \texttt{GINA} cluster at SURFsara, part of the Dutch national e-infrastructure \citep{dutch_einfra}. We test the software performance as a function of several parameters, including averaging parameters, number of CPUs and calibration model size. Additionally, we test the performance of the underlying infrastructure, i.e. queuing  and download time, for the same parameters. Finally, we compare those isolated tests with our production runs of the \texttt{prefactor} pipeline to measure the overhead incurred by running on a shared system. 

We discover that the computationally intensive LOFAR processing steps scale linearly with data size, and calibration model size. Additionally, we find that the time taken by these steps is inversely proportional to the number of CPUs used. We discover that the time to download and extract data on the \texttt{GINA} cluster is linear with size up to 32GB, but becomes slower beyond this data size. We also find that the queuing time on the \texttt{GINA} cluster grows exponentially for jobs requesting more than 8 CPUs. We validate these isolated tests with production runs of LOFAR data from the past six months. We combine all these tests into a single model and show its prediction power by testing the processing time for different combinations of parameters. Finally, we discuss the utility of our method, the results in this work and applications to the SKSP projects, the broader impact of our results to LOFAR processing and the applications for other large astronomical surveys. The major contributions of this work can be summarized as:

\begin{itemize}
    \item A model of processing time for the  LOFAR Direction Independent Calibration Pipeline.
    \item A model of queueing time and file transfer time which is used by current or future jobs processed on the \texttt{GINA} cluster.
    \item Quantification of overheads incurred when processing in production. 
    \item Validation of our methods with discussion of future applications. 
\end{itemize}

We introduce LOFAR processing and other related work in Section \ref{sec:related} and describe our software setup and data processing methods in Section \ref{sec:methods}. We present our results and performance model in Section \ref{sec:results} and discussions and conclusions in Section \ref{sec:discussions}.

\section{Related Work}\label{sec:related}
In previous work, we have parallelized the Direction Independent LOFAR pipeline on a High Throughput infrastructure \citep{mechev17}. While this parallelization has helped accelerate data processing for the SKSP project, creating a performance model of our software is required if we are to predict the resources taken by future jobs. This model will be particularly useful in understanding how processing parameters will affect run time.  

In previous work, we have parallelized the Direction Independent LOFAR pipeline on a High Throughput infrastructure \citep{mechev17}. While this parallelization has helped accelerate data processing for the SKSP project, creating a performance model of our software is required if we are to predict the resources taken by future jobs. This model will be particularly useful in understanding how processing parameters will affect run time.  

Performance modelling on a distributed system is an important field of study related to grid computing. A good model of the performance of tasks in distributed workflows can help more efficiently schedule these jobs on a grid environment \citep{grid_perform_model}. The performance modeling systems require knowledge of the source code and an analytical model of the slowest parts of the code \citep{semi_analytical_model}. Many systems exist to model the performance of distributed jobs \citep{barnes2008regression, semi_analytical_model,performance_prediction,Witt2018PredictivePM}, with some employing Black Box testing \citep{cross_platform_black_box, mapreduce_analysis} or tests on scientific benchmark cases \citep{synthetic_memory_prediction}. Such performance analysis does not require intimate knowledge of the software and can be applied on data obtained from processing on a grid infrastructure.

Empirical modelling is useful in finding performance bugs in parallel code \citep{scalability_bugs} and modelling the performance of big data architectures \citep{mean_field_modeling}. The insights from these models are used to optimize the architecture of the software system or determine bottlenecks in processing. Here, we use empirical modelling to determine how the LOFAR \texttt{prefactor} performance scales with different parameters. 

\section{Processing Setup }\label{sec:methods}

\interfootnotelinepenalty=10000
Using the LOFAR software installation described in \cite{mechev17}, we processed a typical LOFAR SKSP observation\footnote{LOFAR Observation ID L658492, co-ordinates [17h42m21.785, +037d41m46.805] observed by the LOFAR High Band Array for 8 hours between 2018-06-20 and 2018-06-21.}, while changing the averaging rate  in time and frequency. Changing these averaging parameters will change the final data size (with the data sizes studied shown in Table \ref{table:averaging}). We test the processing time for different averaging parameters by running 15 runs per parameter step. 

The data used by the LOFAR surveys is archived at a time resolution of 1 second intervals and frequency resolution of 16 channels per subband (equivalent to 12kHz channel width). While some of the processing steps such as flagging of Radio Frequency Interference and removal of bright off-axis sources produce better results when performed on the high-resolution data, later steps can be performed on averaged data with little impact on the final product quality. To speed up processing, the raw data is averaged in time and frequency, decreasing the input data size to later tasks. The main aims of the LoTSS survey can be accomplished if the final data products from the \texttt{prefactor} pipeline are averaged to a time resolution of 8 seconds per sample and frequency of 2 channels per subband. These averaging parameters correspond to a reduction in data size by a factor of 64. Nevertheless, other science cases require less averaging of the data. Our aim is to understand how the processing of this larger data will scale. To measure the scalability of processing, we measure the performance of the \texttt{prefactor} pipeline for data sizes between the raw data of 64GB/subband and the averaged data of 1GB/subband. The tested data sizes and parameters are shown in table \ref{table:averaging}, and discussed in Section \ref{sec:results_size}.

We performed the scalability tests on a dedicated node of the SURFsara \texttt{GINA} cluster, f18-01. The node is a typical hardware node used by our production LoTSS processing, however it is dedicated for the tests in order to ensure there is no contamination by other software. The node is described in Section \ref{sec:hardware}. 
     
%There are two sources of latency that need to be studied for a true end-to-end model of LOFAR processing. The first is the performance of the LOFAR software on the Dutch grid for a wide range of processing parameters. The second is the overhead, such as job queuing and data movement. Both of these effects depend on similar parameters such as data size and number of CPUs used. 
%We will examine these effects in our study of the LOFAR processing performance by studying the performance at different parameter steps. While some parts of the processing software may change, the infrastructure parts of our performance model can be used independently of the the processing software and can even be applied to other scientific projects running on the \texttt{GINA} cluster. 

\begin{figure}
    \includegraphics[width=0.95\linewidth]{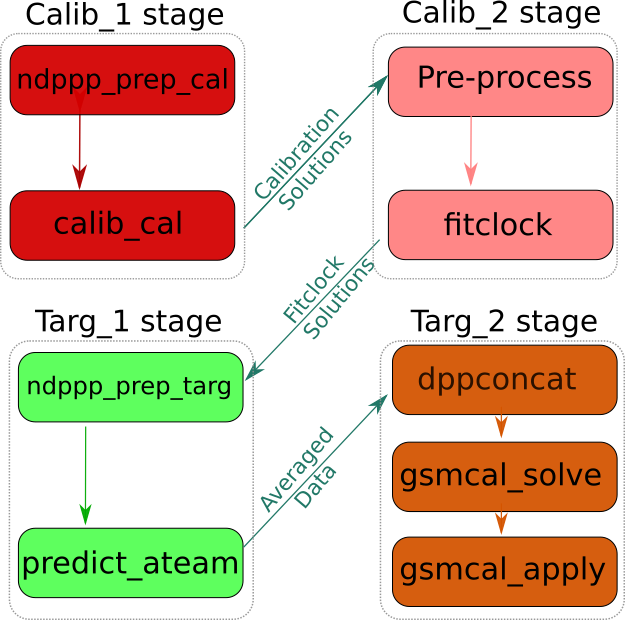}
      \caption{The major steps of the \texttt{prefactor} DI pipeline. }
	\label{fig:prefactor_steps}
\end{figure}

We processed the sample data set with the LOFAR \texttt{prefactor} pipeline. The \texttt{prefactor} version used was the same as we use for the LOFAR SKSP broadband surveys \citep{prefactor_zenodo}. This software consists of several steps executed in sequence, shown graphically in Figure \ref{fig:prefactor_steps}. The important \texttt{prefactor} steps are as follows. The {\fontfamily{qcr}\selectfont predict\_ateam} and {\fontfamily{qcr}\selectfont ateamcliptar} steps predict the contamination by bright off-axis sources and remove these effects respectively. The {\fontfamily{qcr}\selectfont dpppconcat} step is responsible for concatenating 10 subbands into a single file which is in turn calibrated. The step {\fontfamily{qcr}\selectfont gsmcal\_solve} is responsible for calibration of the data against a model of the radio sky. The solutions produced by {\fontfamily{qcr}\selectfont gsmcal\_solve} are used by {\fontfamily{qcr}\selectfont gsmcal\_apply} and applied to the scientific observation.

\subsection{Processing Metrics}
The goal for our scalability model is to understand the effect of several parameters on the job completion time of LOFAR software. We do this by testing the processing time for various values of data size, number of CPUs used and sky model size.

\begin{table}[!ht]
\centering
\begin{tabular}{||c| c | c | c||} 
 \hline
\multicolumn{1}{|p{1.7cm}|}{\centering Averaging \\ ratio} & \multicolumn{1}{|p{1.8cm}|}{\centering Time averaging \\ parameter (sec)} &  \multicolumn{1}{|p{1.7cm}|}{\centering Channels per Subband} & \multicolumn{1}{|p{1.8cm}|}{\centering Averaged \\ Size (Gb) }\\ [0.5ex]
 \hline
 \rowcolor{Gray}
  \hline
 64x & 8   & 2   &  1.235   \\ 
  \hline
 32x & 4   & 2   &  2.459   \\ 
 16x & 2   & 2   &  4.906   \\ 
 8x & 1   & 2   &  9.802   \\ 
 4x & 1   & 4   &  18.00  \\ 
 2x & 1   & 8   &  36.72  \\ 
 1x & 1   & 16   &  66.88  \\[1ex] 
 \hline
\end{tabular}
\caption{Averaging parameters and final data sizes tested for the sample LOFAR SKSP observation. The raw data is 64 GB per subband. The LOFAR SKSP data processing uses averaging parameters of 8 seconds and 2 channels per subband. This reduces the raw data by a factor of 64. We highlight the data size used in the LOFAR SKSP survey.   }
\label{table:averaging}
\end{table}

The slowest step of the \texttt{prefactor} pipeline is the {\fontfamily{qcr}\selectfont gsmcal\_solve} step, which performs the gain calibration against a model of the radio sky. This step operates on a concatenated data set that consists of 10 subbands. We obtain the calibration model through the TGSS sky model creator\footnote{Accessible at \href{http://tgssadr.strw.leidenuniv.nl/doku.php}{the TGSS ADR portal}.}. By default this service creates a text file describing the sky-model from the TGSS survey \citep{tgssadr} as a combination of gaussian and point sources. By default, it sets a threshold of sources brighter than 0.3 Jy. 
Lowering this threshold creates longer sky-model files with more faint sources, while increasing it will return only the few brightest sources. Since sky model calibration requires converting the sky-model into modelled visibilities \citep[e.g.][]{dppp, radio_visibility_sage,app_synth}, a longer sky model will increase the time taken to gain calibrate a data set. We created 7 sky models with a flux cutoff ranging between 0.05 Jy and 1.5 Jy. The number of sources in the resulting models are listed in Table \ref{table:skymodels}. 
For production\footnote{The query used to obtain model 3 is at the following link \url{http://bit.ly/tgss_model}}, we used the minimum sensitivity parameters for model 3.
%%%%%Each line of these model files corresponds to one source, modelled either as a point or an ellipse), hence the second column also lists the number of sources per sky model file. %Suggested remove by Huub

\begin{figure}
    \includegraphics[width=0.95\linewidth]{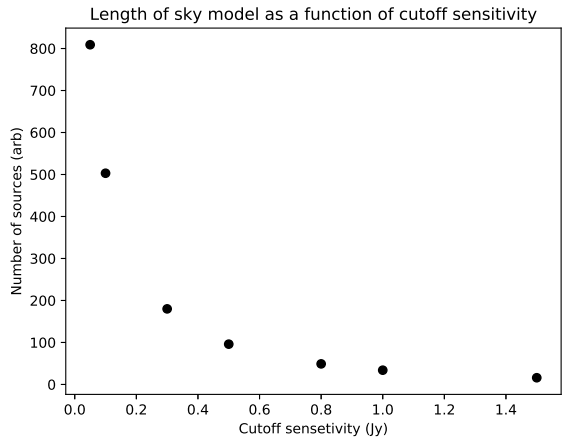}
      \caption{The size of the sky model (measured in number of sources) increases exponentially as we decrease the flux cutoff of the model (i.e. increase the sensitivity).}
	\label{fig:skymodel_size}
\end{figure}

It is important to note that the complexity and accuracy of the sky model depend on the direction of observation and the ionospheric conditions in which the observation was performed. As such, our test is only a heuristic for predicting the run-time based on the calibration model length. Additionally, it is notable that the number of sources is exponentially dependent on the minimum sky model sensitivity (seen in Figure \ref{fig:skymodel_size}, more in \citealt{tgssadr,Wendy_bootes}). According to this relationship, even a modest decrease in sensitivity cutoff can significantly decrease the size of the model.

\begin{table}[!ht]
\centering
\begin{tabular}{||c| c | c||} 
 \hline
 Sky model \# & min sensitivity & \# sources  \\ [0.5ex] 
 \hline
 model 1 & 0.05 Jy & 809    \\ 
 model 2 & 0.1 Jy & 503   \\
 \rowcolor{Gray}
  \hline
 model 3 & 0.3 Jy & 180   \\
  \hline
 model 4 & 0.5 Jy & 96  \\
 model 5 & 0.8 Jy & 49   \\ 
 model 6 & 1.0 Jy & 34   \\
 model 7 & 1.5 Jy & 16   \\[1ex] 
 \hline
\end{tabular}
\caption{List of test sky models. Model 3 is created with the parameters used in our production processing of LOFAR data. All models include objects within 5 degrees from the centre of the pointing.  }
\label{table:skymodels}
\end{table}

Finally, the number of CPU cores (henceforth just `CPUs') used by each step is a parameter that can be optimized for the entire pipeline. While increasing the number of CPUs can make some steps run faster, requesting jobs that reserve a large number of CPUs will take longer to launch on shared infrastructure. In order to understand the interplay between these effects, we study the queuing time and processing time as a function of the number of CPUs. For the parameter steps we choose to test 1, 2, 3, 4, 8 and 16 CPUs. 

\subsection{Infrastructure Performance}
%%%%%Staging files at different locations can be modelled using historical data, discuss the importance of knowing the time it takes to stage data and the prediction of future performance. 

In production, we run hundreds of LoTSS jobs on a cluster supporting several different use cases. The requested resources on this cluster are allocated by a job queue, in our case implemented by the glite workload management system \citep{glite-wms}. As queuing jobs can take a significant amount of time, we test the queuing time as a function of the number of requested CPUs. In order to do that, we create test jobs that log the launch time and submit them, requesting 1, 2, 3, 4, 8 and 16 CPUs. We run 10 to 15 tests for each parameter step to ensure that we capture system variability at different times of day during the week and the weekend. 

Besides queuing, time is also spent during downloading and unpacking data, as well as packing and uploading the results. Despite using no compression to pack the data, untarring and tarring large files still takes time depending on the system workload. We measure the time taken to transfer and unpack data of different sizes. The data sizes we chose were 0.5GB, 1GB, 2GB, 4GB, 8GB, 16GB, 32GB and 64GB. As our largest data sets are 64GB and our smallest results are $\sim$0.2GB, these values span a realistic range expected for LOFAR data processing. We test this by uploading mock data to the dCache storage pool at SURFsara and launching a small 1 CPU job on the cluster, which downloads and untars the data and logs the start time of each step. We present the results of this test in the next section.

\subsubsection{Software Versions}\label{sec:software_versions}
For the current test, we use the LOFAR software stack, version 2.20.2 \citep{cookbook}. This software was compiled on a virtual machine and distributed using the CERN CVMFS virtually mounted file system \citep{cvmfs2008}. We use this software version and distribution method as it is the same software version and distribution used to process the data for the LOTSS Data Release 1. 

\subsection{Test Hardware}\label{sec:hardware}

We test the LOFAR software on a reserved node on the SURFsara \texttt{GINA} cluster. The node, f18-01 has 348 GB of RAM, 3TB of scratch space\footnote{More detailed specifications are at \href{http://docs.surfsaralabs.nl/projects/grid/en/latest/Pages/Service/system_specifications/gina_specs.html}{the \texttt{GINA} specification page linked here}}. The CPU is an Intel Xeon(R) Gold 6148 CPU with 40 cores clocked at 2.40GHz.  As this hardware node was reserved, there was no other scientific processing aside from our tests, meaning there was no resource contention aside for that inherent in the LOFAR software. In the results section, we compare these isolated runs with processing results over the past two years. 

\section{Results}\label{sec:results}

Using a test data set, we tested the LOFAR \texttt{prefactor} target pipeline on the SURFsara \texttt{GINA} cluster. First we will present the tests done in an isolated environment and then compare them to the run time in production on a shared infrastructure. We will integrate all the results in a complete model which can be used to predict processing time for a variety of parameters. Finally, we will make some predictions on the run time of our processing based on the model and validate these predictions. 

Since we are  processing a sample data set in the context of the LOFAR Surveys project, we will compare these tests with the production runs of our pipeline. In production, we run  the {\fontfamily{qcr}\selectfont gsmcal\_solve} step with a data size of 1GB, a sky model with 180 sources and 8 CPUs.

\subsection{Isolated Environment tests}
We first tested the LOFAR software in isolation in order to determine the scalability of processing time in terms of data size. We run the entire \texttt{prefactor} target pipeline which which removes Direction Independent Calibration errors from a LOFAR science target. In the following sections, we present the models obtained from these tests.  

\subsubsection{Input Data Size}\label{sec:results_size}
\begin{figure*}[t!]
        \centering
        \begin{subfigure}[b]{0.44\textwidth}
            \centering
            \includegraphics[width=\textwidth]{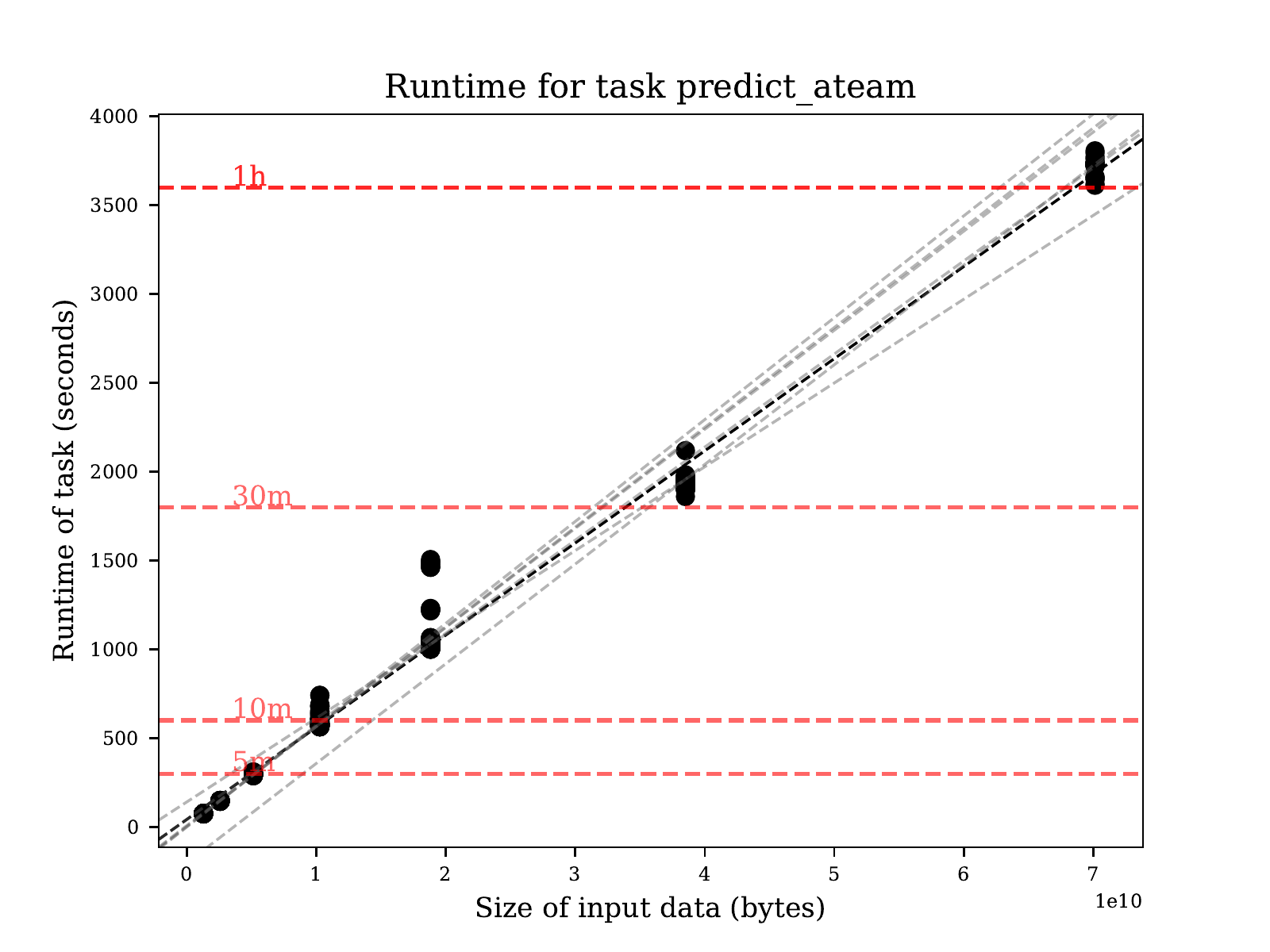}
            \caption[]%
            {{\small Tests of the  {\fontfamily{qcr}\selectfont predict\_ateam} step for input data size ranging from 1GB to 64 GB. This step calculates the contamination from bright off-axis sources. Dashed lines are shown connecting each pair of points, to highlight the trend. We can see that the data can be described well by a linear model. We show the model in Equation \ref{eq:predictateam} in black.}}   
            \label{fig:predict_ateam}
        \end{subfigure}
        \hfill
        \begin{subfigure}[b]{0.44\textwidth}  
            \centering 
            \includegraphics[width=\textwidth]{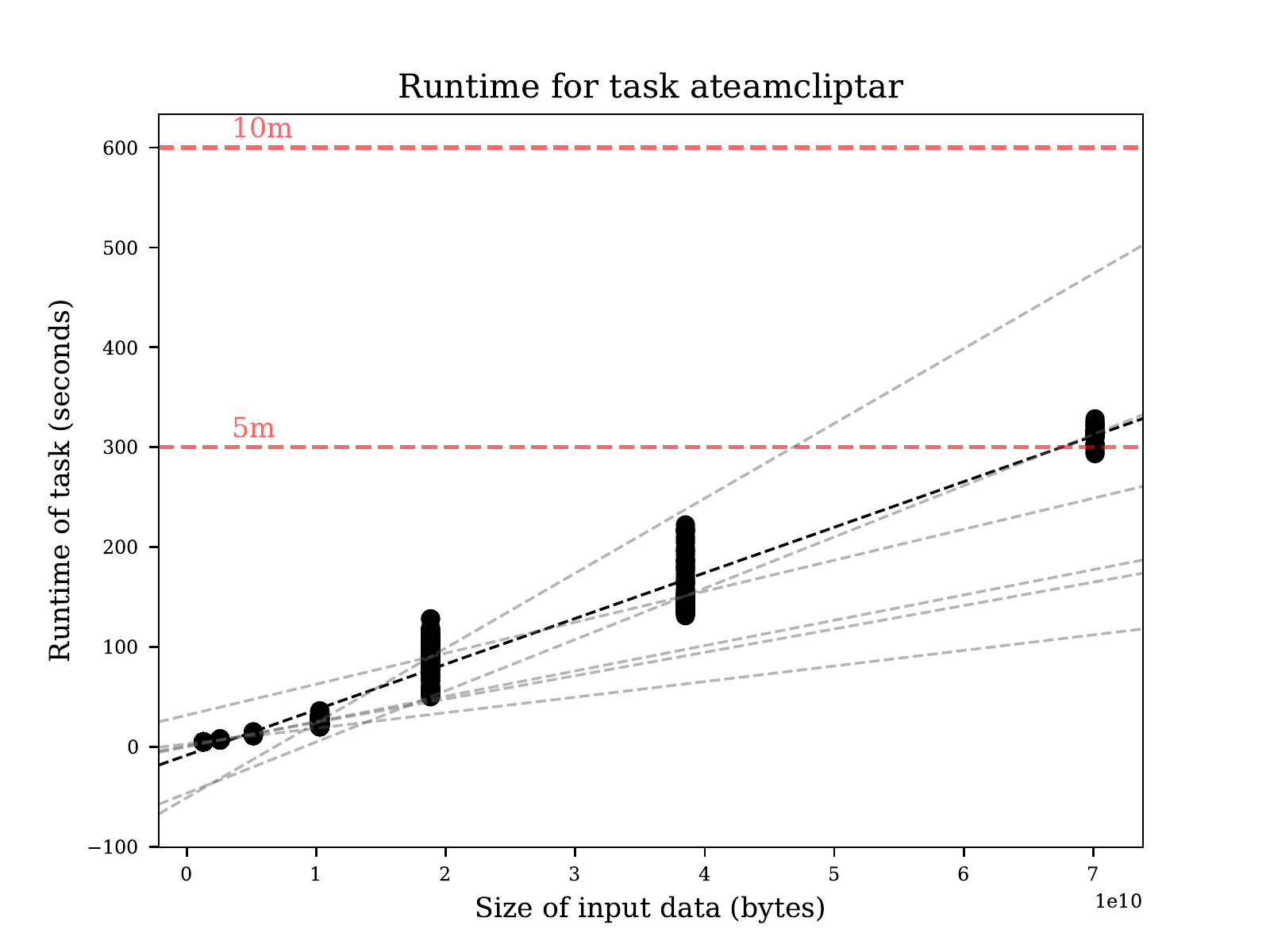}
            \caption[]%
            {{\small Tests of the {\fontfamily{qcr}\selectfont ateamcliptar} step for input data size ranging from 1GB to 64 GB. This step removes the contamination from bright off-axis sources. We can see that the data fits a linear model. We show the model in Equation \ref{eq:ateamcliptar} in black.}}    
            \label{fig:ateamcliptar}
            \end{subfigure}
        \vskip\baselineskip
        \begin{subfigure}[b]{0.44\textwidth}   
            \centering 
            \includegraphics[width=\textwidth]{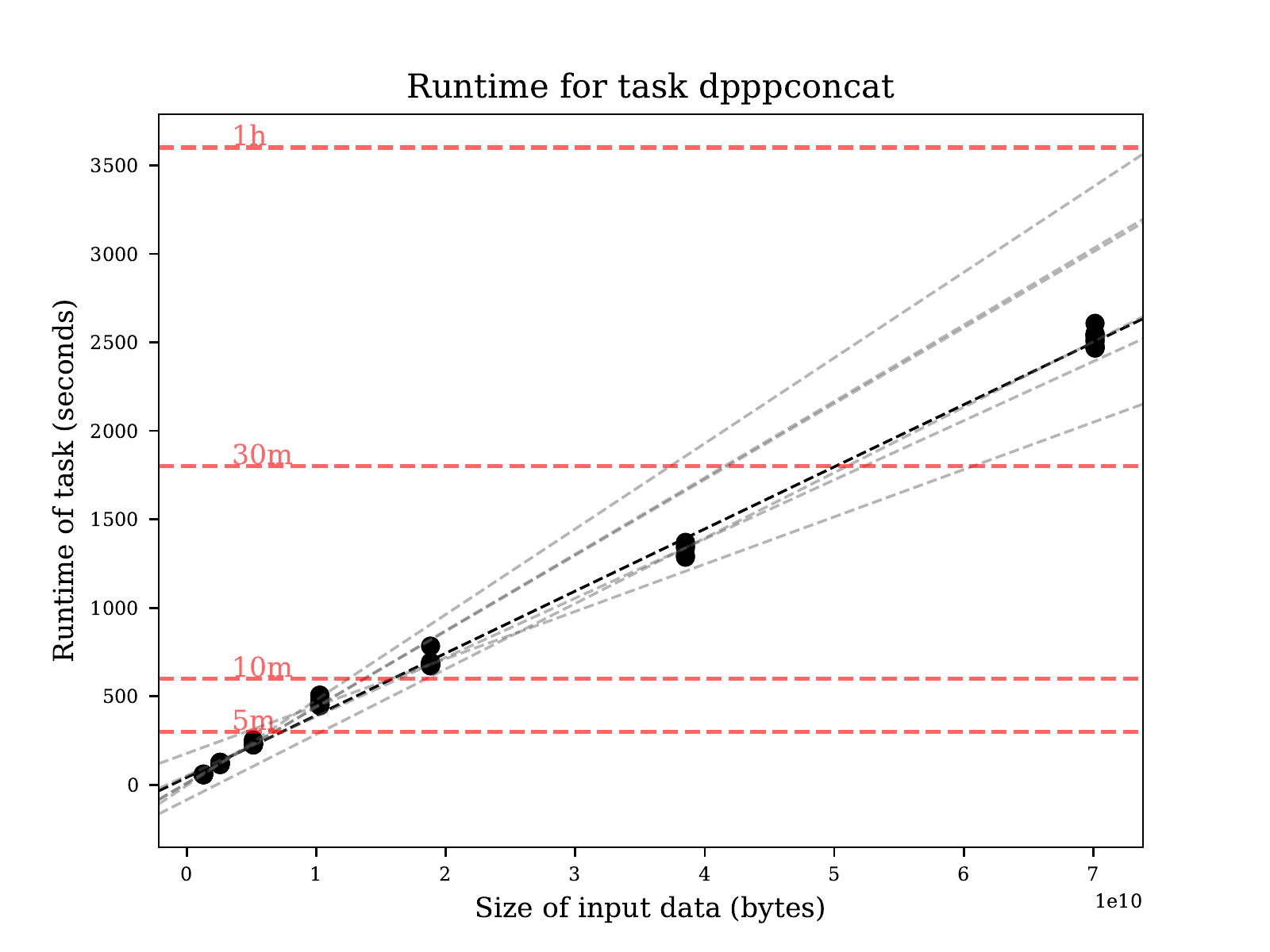}
            \caption[]%
            {{\small Tests of the {\fontfamily{qcr}\selectfont dpppconcat} step for input data size ranging from 1GB to 64 GB. This step concatenates 10 files into a single measurement set.  We can see that the data fits a linear model. We show the model in Equation \ref{eq:dpppconcat}  in black.}}    
            \label{fig:dpppconcat_size}
        \end{subfigure}
        \hfill
        \begin{subfigure}[b]{0.44\textwidth}   
            \centering 
            \includegraphics[width=\textwidth]{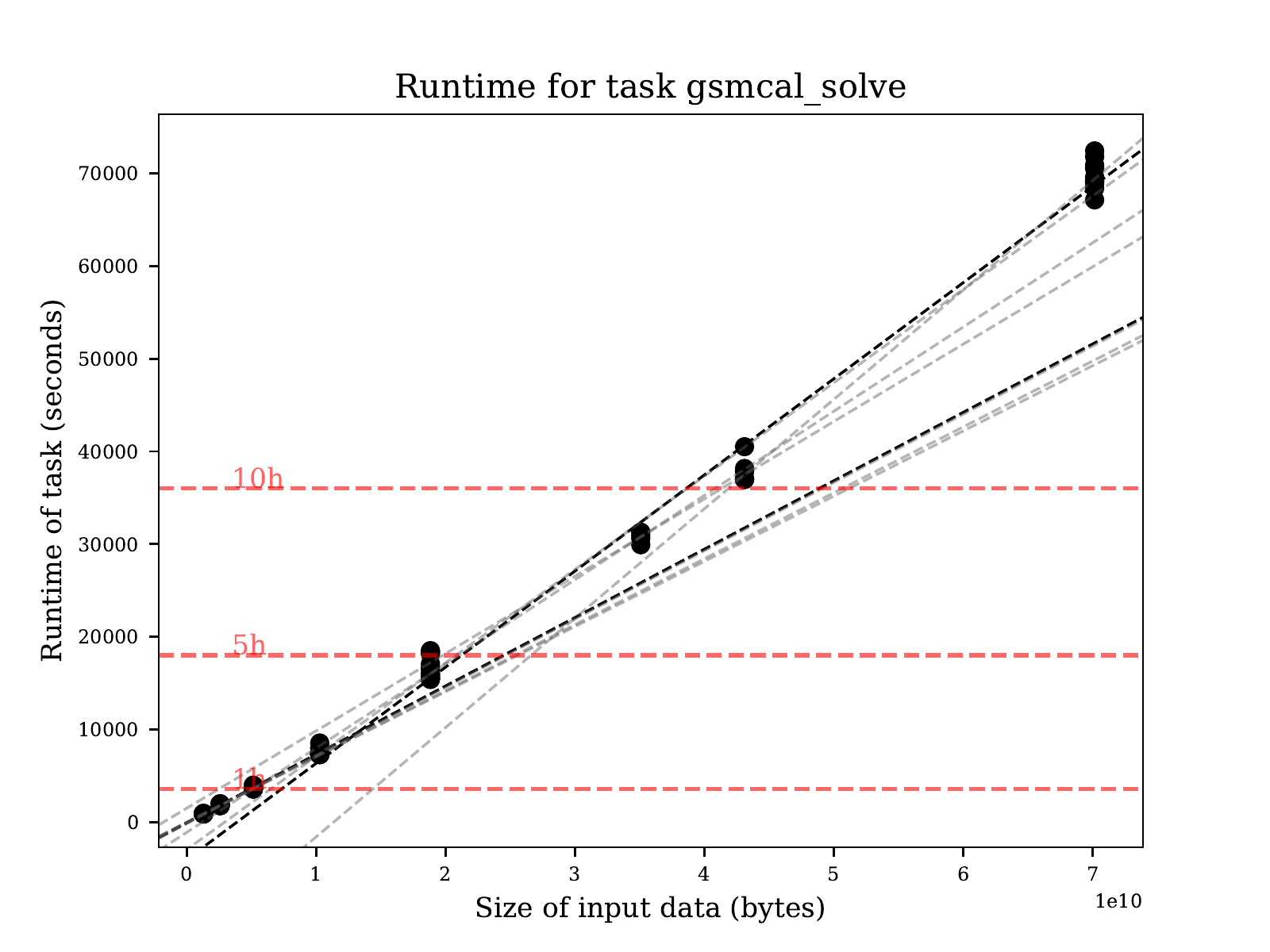}
            \caption[]%
            {{\small Tests of the {\fontfamily{qcr}\selectfont gsmcal\_solve} step for input data size ranging from 1GB to 64 GB. This step performs gain calibration of the concatenated data set against a sky model. It is the slowest and most computationally expensive \texttt{prefactor} step. We fit two linear models, for data below 16GB and above 16GB. We can see the two models, shown in  (Equation \ref{eq:gsmcalsolve}) as two black dashed lines, intersecting at 1GB.}}    
            \label{fig:gsmcalsolve_size}
        \end{subfigure}
        \caption[ ]
        {\small Plots of the run time as a function of input data size} 
\end{figure*}

\begin{figure}
\includegraphics[width=0.95\linewidth]{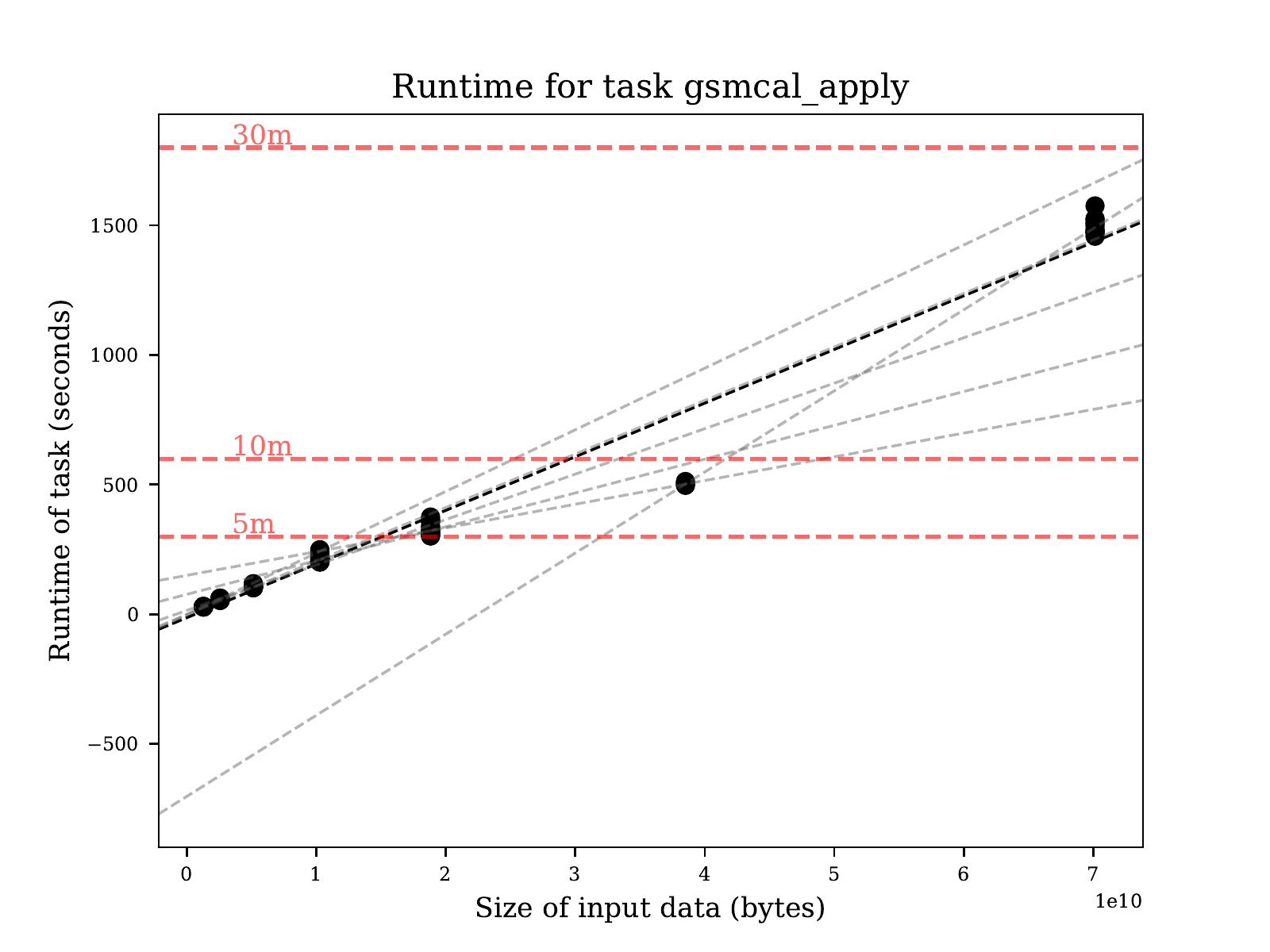}
        \caption{\small Tests of the {\fontfamily{qcr}\selectfont gsmcal\_apply} step for input data size ranging from 1GB to 64 GB. This step applies the calibration solutions to the data. We can see that the data fits a linear model, described in Equation\ref{eq:gsmcalapply}, as the black dashed line.}
            \label{fig:gsmcalapply_size}
\end{figure}

LOFAR data can be averaged to different sizes based on the scientific requirements. Smaller data sets are processed faster, so it is important to understand the effect of data size on processing time as measured by wall-clock time. We show the processing time for our test data set, averaged to different sizes for several \texttt{prefactor} steps in Figures \ref{fig:predict_ateam}- \ref{fig:gsmcalsolve_size} and \ref{fig:gsmcalapply_size}. We run this test using 8 CPUs. The figures also show linear fits for consecutive pairs of parameter steps, in gray dashed lines, used to help guide the selection of parametric model. 

All of the steps show a linear behavior with respect to input data size, while the {\fontfamily{qcr}\selectfont gsmcal\_solve step} is best fit by two linear relationships, for data smaller and larger than 16 GB. The linear fit to the run times are shown in Equations \ref{eq:predictateam}-\ref{eq:gsmcalapply}. The equations show the processing time as a function of the data size ($\mathcal{S}$), with the slope in the units of seconds/byte. The fits are also shown in Figures \ref{fig:predict_ateam} to \ref{fig:gsmcalapply_size} as a black dashed line.

\begin{equ*}
\begin{subequations}
\begin{align}
        T_{predict\_ateam}=5.19\times10^{-8}\mathcal{S}+4.20\times10^1 \label{eq:predictateam} \\
        T_{ateamcliptar}=4.57\times10^{-9}\mathcal{S}-8.42\times10^0 \label{eq:ateamcliptar} \\
        T_{dpppconcat}=3.51\times10^{-8}\mathcal{S}+4.20\times10^1 \label{eq:dpppconcat} \\
        T_{gsmcal\_solve}=\begin{cases}
                          7.38\times10^{-7}\mathcal{S}-8.20\times10^1 &|\mathcal{S}\leq1.6\times10^{10}\\
                          1.04\times10^{-6}\mathcal{S}-4.04\times10^3 & |\mathcal{S}>1.6\times10^{10}
    \end{cases} \label{eq:gsmcalsolve} \\
        T_{gsmcal\_apply}=2.07\times10^{-8}\mathcal{S}-1.38\times10^1 \label{eq:gsmcalapply}    
\end{align}
\label{eq:runtime_size_models}
\end{subequations}
\caption{Equations describing the processing time of five \texttt{prefactor} steps as a function of the input data size ($\mathcal{S}$) in bytes.}
\end{equ*}

\subsubsection{Calibration Model Size}
To test the effect of the calibration model size on run time, we test our calibration with several different lengths of the sky model file. We create these models by changing the minimum sensitivity using values ranging from 0.05 Jy to 1.5 Jy. The most sensitive model (0.05 Jy) had 809 sources while the 1.5 Jy model had only 16 sources. 

\begin{figure}
    \includegraphics[width=0.95\linewidth]{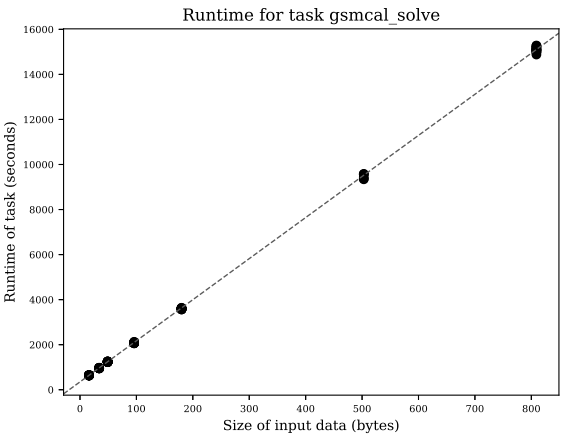}
      \caption{The processing time of the {\fontfamily{qcr}\selectfont gsmcal\_solve} step is linear with the size of the sky model as measured by the number of sources.}
	\label{fig:skymodel_run_lenght}
\end{figure}

\begin{figure}
    \includegraphics[width=0.95\linewidth]{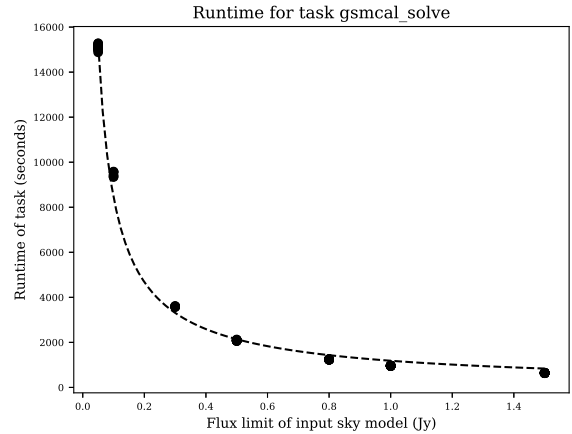}
      \caption{The run time of the {\fontfamily{qcr}\selectfont gsmcal\_solve} step as a function of the cutoff sensitivity is not linear. As shown in Figure \ref{fig:skymodel_size}, the number of sources increases exponentially as the minimum sensitivity decreases. The dashed line shows the model fitted in Equation \ref{eq:skymodel_flux}. }
	\label{fig:skymodel_run_sens}
\end{figure}

Figure \ref{fig:skymodel_run_lenght} shows that the calibration time is directly proportional to the length of the sky model. Figure \ref{fig:skymodel_run_sens} shows the run time as a function of the processing parameter: the cutoff sensitivity. As the relationship between the number of sources and cutoff sensitivity is a power law, here we see the same relationship holding for processing time.

We model the processing time as a function of the cutoff frequency using a power law, and fit the data to the function $y=\alpha\cdot \mathcal{F}^{-k}$. Our fit found the best model to be shown in Equation \ref{eq:skymodel_flux}, where $\mathcal{F}$ value is the cutoff flux in Jansky and $T$ is the run time in seconds. 

\begin{figure*}
    \includegraphics[width=0.95\linewidth]{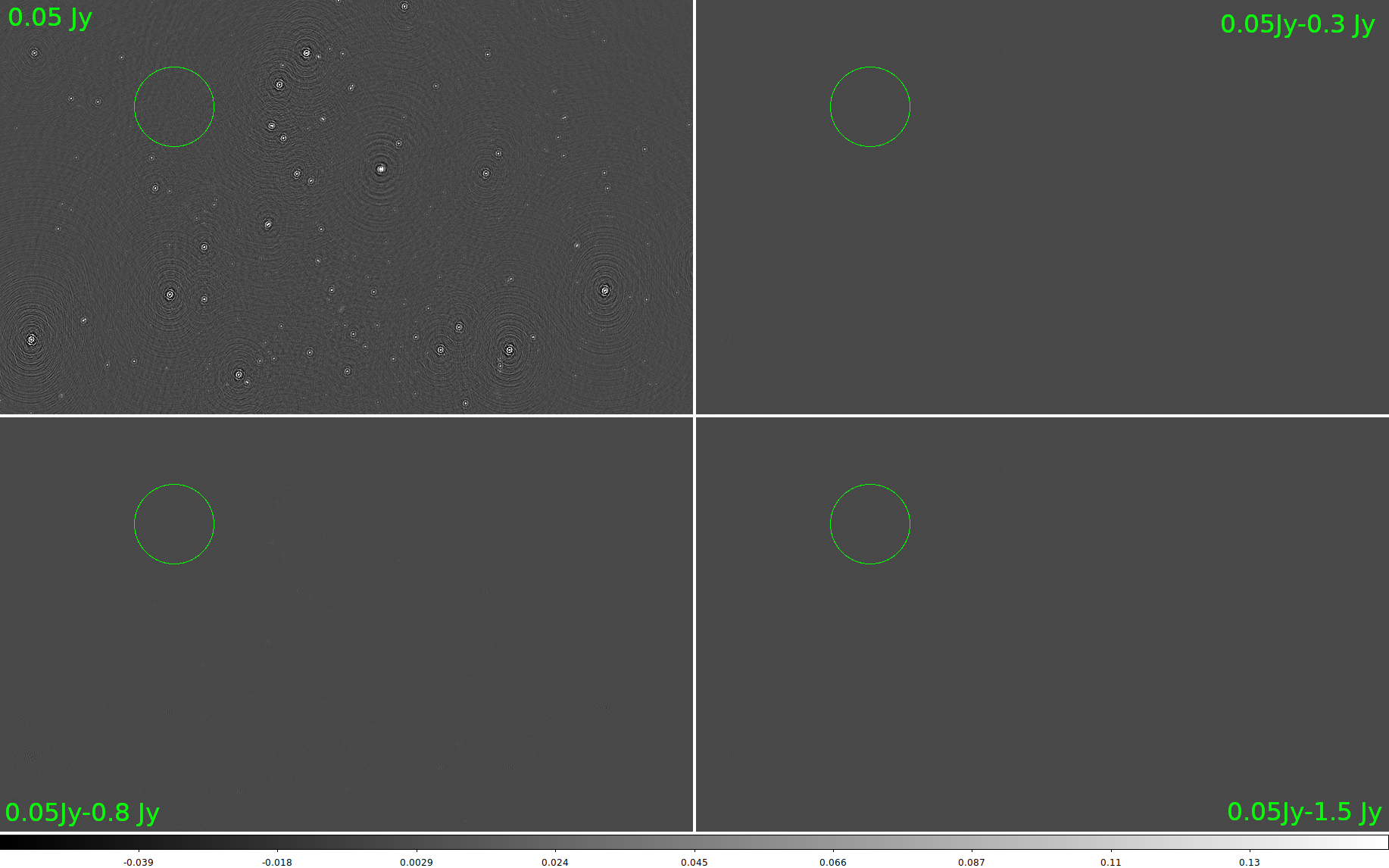}
      \caption{Four images made using the \texttt{wsclean} software \citep{wsclean} from the data set\protect\footnotemark. The four images were calibrated with sky models of various flux cutoffs ranging from 0.05Jy (top left) to 1.5Jy (bottom right). Flux statistics for the green regions in the four images are listed in Table \ref{table:skymodel_RMS}. The top right, and bottom two quadrants show the pixel difference between the 0.05Jy image and the 0.3Jy, 0.8Jy and 1.5Jy images respectively. The four images are all on the same scaleThe green region shows the same area in all four quadrants.  }
	\label{fig:skymodel_images}
\end{figure*}

We show four images made from data sets in Figure \ref{fig:skymodel_images}. The top left image is calibrated with a 0.05Jy and the other three show the difference between the top left image and the images created from the 0.3Jy, 0.8 Jy and 1.5 Jy data. The statistics for the four images, taken from the regions in green on Figure \ref{fig:skymodel_images}) are shown in Table \ref{table:skymodel_RMS}. We discuss the implication and caveats of these results in Section \ref{sec:discussions}.

\begin{table}[h!]
\centering
\begin{tabular}{||p{2.8cm}| c | c ||} 
 \hline
 Calibration Model Flux Cutoff & \# of sources& RMS Noise (Jy) \\ %%& std dev (Jy) \\ [0.5ex]
 \hline
 0.05Jy & 809 &0.00402834   \\ %& 0.004026    \\ 
  \rowcolor{Gray}
  \hline
 0.3 Jy & 180 &0.00402311 \\ %& 0.004020 \\
 \hline
 0.8 Jy & 49 &0.00404181 \\ %& 0.004039 \\  
 1.5 Jy & 16 &0.00410204 \\ %& 0.004105\\
 \hline
\end{tabular}
\caption{Statistics for an empty region for the four images shown in Figure \ref{fig:skymodel_images}. The 0.3Jy model, here shown shaded in gray,  is the one used in production.  }
\label{table:skymodel_RMS}
\end{table}

\begin{equ}
\begin{equation}
    T=1185\cdot \mathcal{F}^{-0.854}
\label{eq:skymodel_flux}
\end{equation}
\caption{Processing time for the {\fontfamily{qcr}\selectfont gsmcal\_solve} step as a function of the flux cutoff of the calibration model ($\mathcal{F}$) in Jansky}
\end{equ}

\footnotetext{Using the command wsclean -absmem 50 -niter 3 -size 4096 4096 -scale 20asec}

\subsubsection{Number of CPUs}
\begin{figure}
    \includegraphics[width=0.95\linewidth]{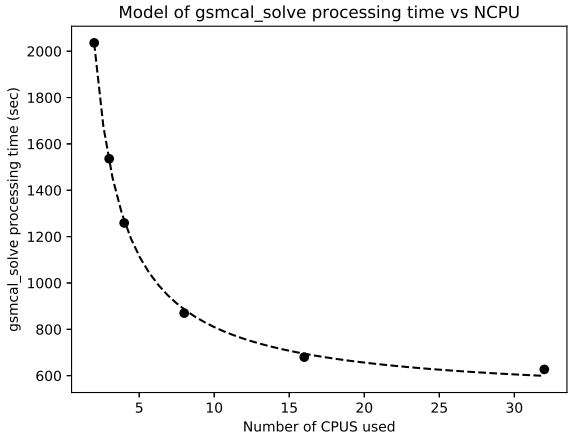}
      \caption{The processing time of the {\fontfamily{qcr}\selectfont gsmcal\_solve} step decreases exponentially with the number of CPUs requested. The model in Equation \ref{eq:gsmcal_NCPU} is shown in a dashed line. As this is a 1/x model, it shows diminishing returns past 16 CPUs. }
	\label{fig:gsmcal_solve_NCPU}
\end{figure}

\begin{figure}
    \includegraphics[width=0.95\linewidth]{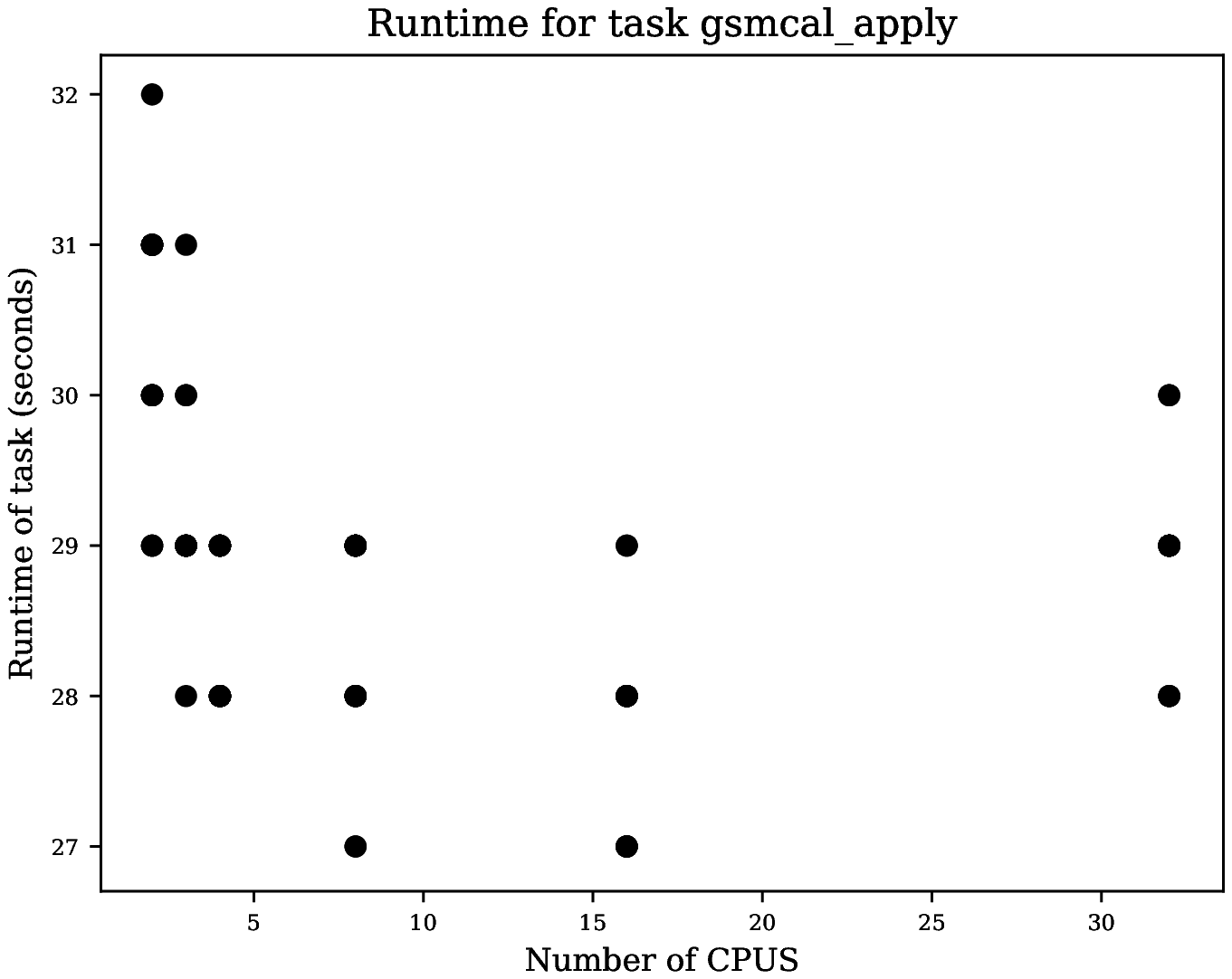}
      \caption{The step that applies the calibration solutions, {\fontfamily{qcr}\selectfont gsmcal\_apply}, does not show a speedup when run on multiple cores, as all runs take roughly 30 seconds to complete.  }
	\label{fig:gsmcal_apply_NCPU}
\end{figure}

One further parameter that can be optimized is the number of CPUs requested when the job is launched. We investigated the processing speedup as a function of the number of CPUs for the \texttt{prefactor} target pipeline. From the steps tested, only the {\fontfamily{qcr}\selectfont gsmcal\_solve} step shows a significant speedup as the number of CPUs is increased. The run time of this step is an inverse power law with respect to the number of CPUs as seen in Figure \ref{fig:gsmcal_solve_NCPU}. Unlike the solving step, the step applying the calibration solutions ({\fontfamily{qcr}\selectfont gsmcal\_apply}) is constant in time with respect to the number of CPUs as seen in Figure \ref{fig:gsmcal_apply_NCPU}. The difference in performance is most likely because the {\fontfamily{qcr}\selectfont gsmcal\_apply} code uses a parallel for loop to calculate antenna gains while {\fontfamily{qcr}\selectfont gsmcal\_apply} does not. 

We fit a model with processing time inversely proportional to the number of CPUs used. We show the resulting model in Equation \ref{eq:gsmcal_NCPU}, with the parameter $(\mathcal{N})$ being the number of CPUs used. 

\begin{equ}
\begin{equation}
    T=503.37+\frac{3062.6}{\mathcal{N}}
\label{eq:gsmcal_NCPU}
\end{equation}
\caption{Processing time for the {\fontfamily{qcr}\selectfont gsmcal\_solve} step as a function of ($\mathcal{N}$), the Number of CPUs used by the process.}
\end{equ}

\subsection{Queuing Tests}

\begin{figure}
    \includegraphics[width=0.95\linewidth]{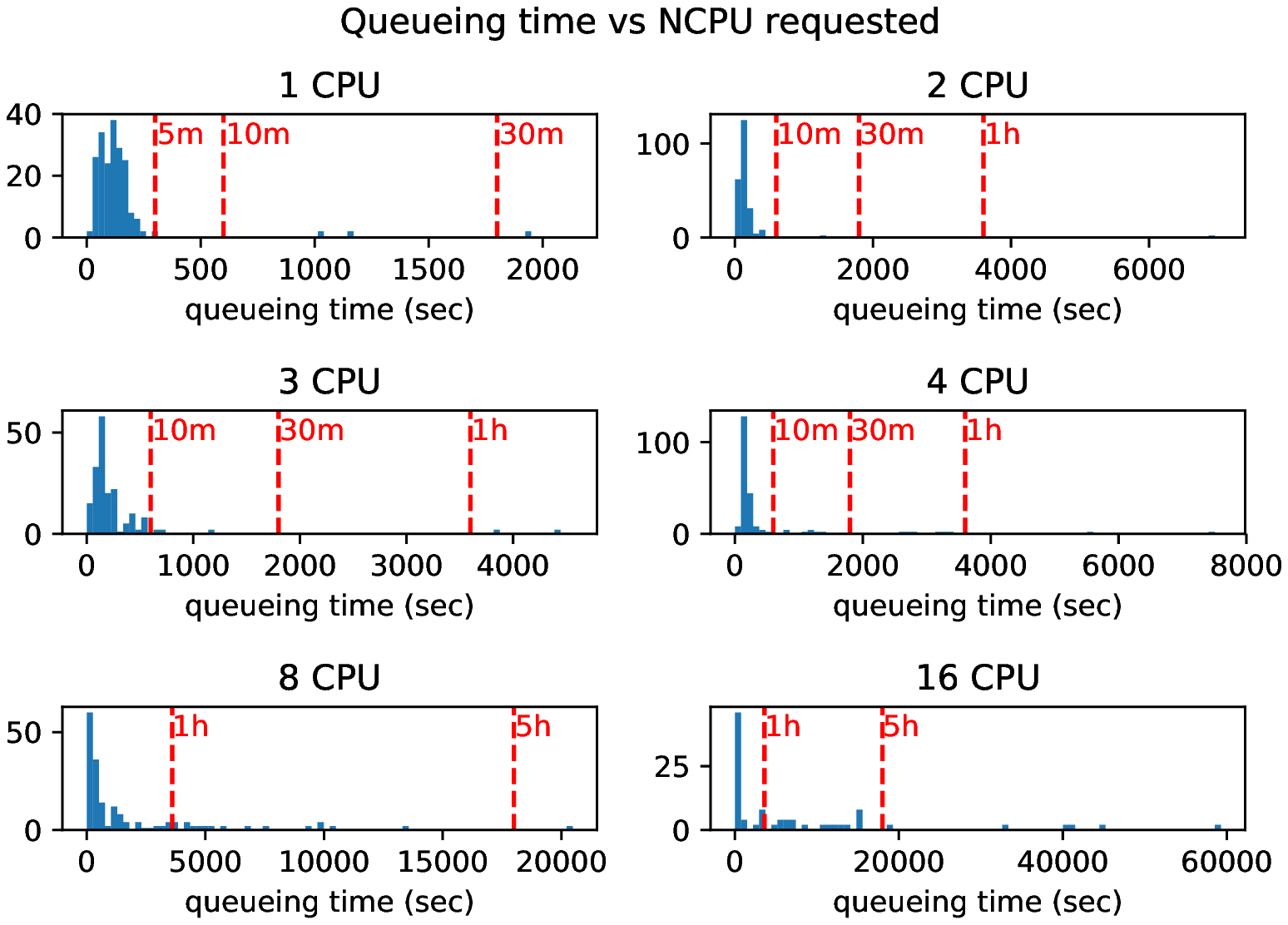}
      \caption{Test randomly submitting jobs to the \texttt{GINA} with different number of requested CPUs. The long tail for 8 and 16 CPU jobs shows that some jobs can take several hours to launch.  }
      
      %Outliers due to grid downtime have been removed \textbf{Actually do this though} }
	\label{fig:queue_NCPU}
\end{figure}

Aside from performance of the LOFAR software, we measured the queuing time at the \texttt{GINA} cluster, as a function of the number of CPUs requested. This data was obtained between 16 Nov 2018 and 10 Dec 2018 for 1,  2, 3 ,4, 8, and 16 CPUs per job. A histogram of the queuing time for these jobs is shown in Figure \ref{fig:queue_NCPU}. Statistics for these runs are in Table \ref{table:queueing_stats}. We use the 75th percentile of the queuing time for each parameter step to fit a model. This scenario will include 75\% of runs and is a good trade-off between ignoring and including outliers. 

\begin{figure}
    \includegraphics[width=0.95\linewidth]{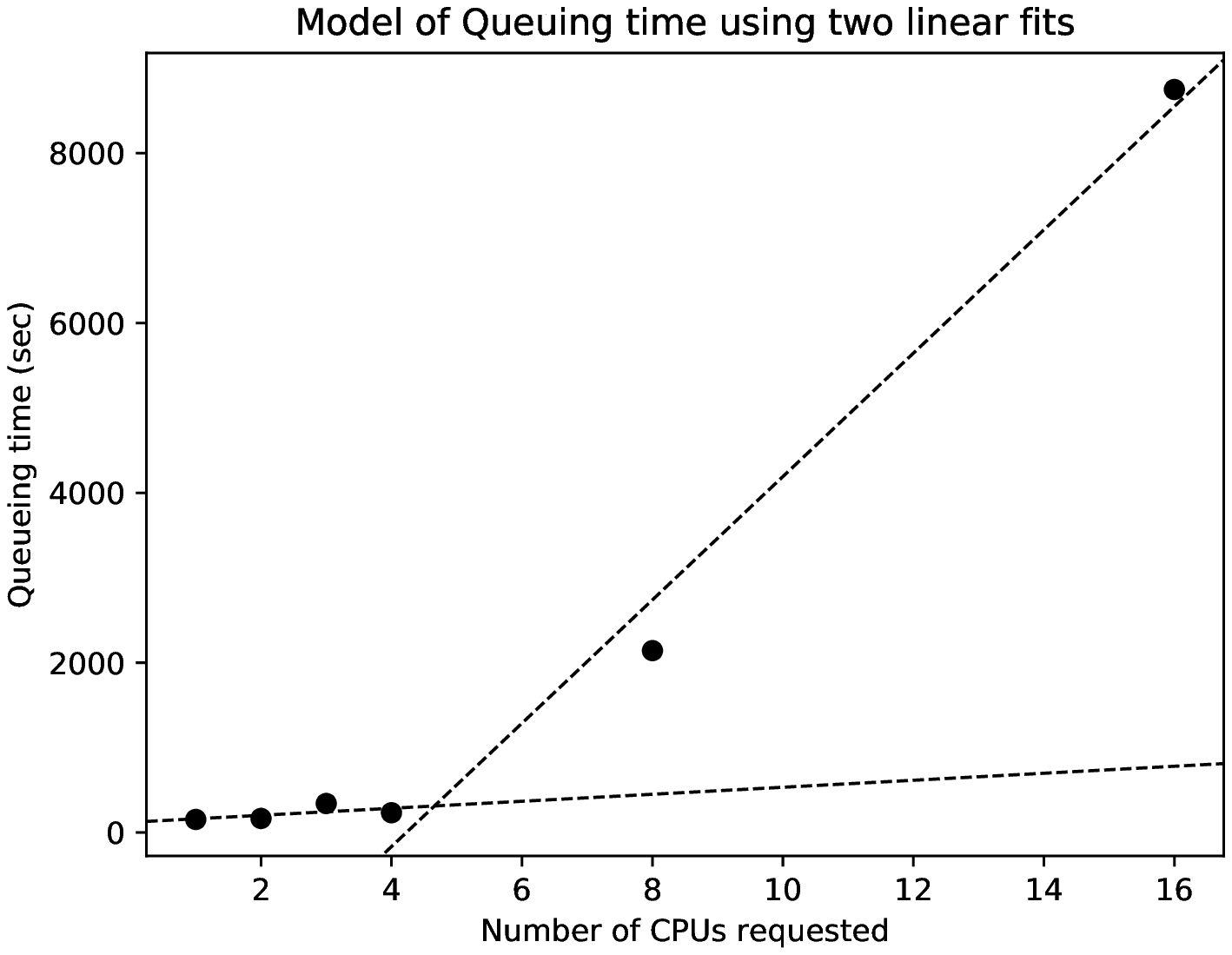}
      \caption{The queuing model built from two linear fits to the queuing times. We use the 75th percentile of the queuing data as a upper bound of job queuing. }
	\label{fig:queue_model}
\end{figure}

We fit two linear models for this queueing time. One model for 1-4 CPUs and one for 4-16 CPUs. The model, as a function of the number of CPUs $\mathcal{N}$ is in equation \ref{eq:queue_model}. The two models are plotted against the 75th percentile of the queuing times (last column in Table \ref{table:queueing_stats}) in Figure \ref{fig:queue_model}.

\begin{table*}[t]
\centering
\begin{tabular}{||p{2.8cm}|c | c | c||} 
 \hline
 NCPU requested & Mean time (sec) & Median time (sec) & 75th percentile (sec)\\ [0.5ex]
 \hline
 1 CPU & 150.5   & 116.2 & 154.1   \\ 
 2 CPU & 201.1 & 125.8 & 165.8 \\
 3 CPU & 296.2 & 152.0 & 243.0 \\
 4 CPU & 498.9 & 167.7 & 233.7\\
 8 CPU & 1944.2 & 428.4 & 2142.4\\
 16 CPU & 7079.0 & 696.4 & 8750.6\\
 \hline
\end{tabular}
\caption{Statistics for queuing time for different values of CPUs requested. Queueing time for jobs requesting less than 8 CPUs is typically less than five minutes. It can drastically increase for larger jobs. }
\label{table:queueing_stats}
\end{table*}

\begin{equ}
\begin{equation}
  T = \begin{cases}
    49.3\cdot\mathcal{N}+ 120 &|\mathcal{N}\leq4\\
    726\cdot\mathcal{N}-3071 & |\mathcal{N}>4
    \end{cases}
  \label{eq:queue_model}
\end{equation}
\caption{The model for the Queuing time as described by two linear models. }
\end{equ}

\subsection{Transfer and Unpacking Time}\label{sec:results_dl}

We tested the downloading and unpacking time for data sizes ranging from 512MB to 64GB. We discovered that the unpacking of files below 64GB scaled linearly with file size, however unpacking individual data sets larger than 16GB becomes considerably slower than downloading it. 

\begin{figure}
    \includegraphics[width=0.95\linewidth]{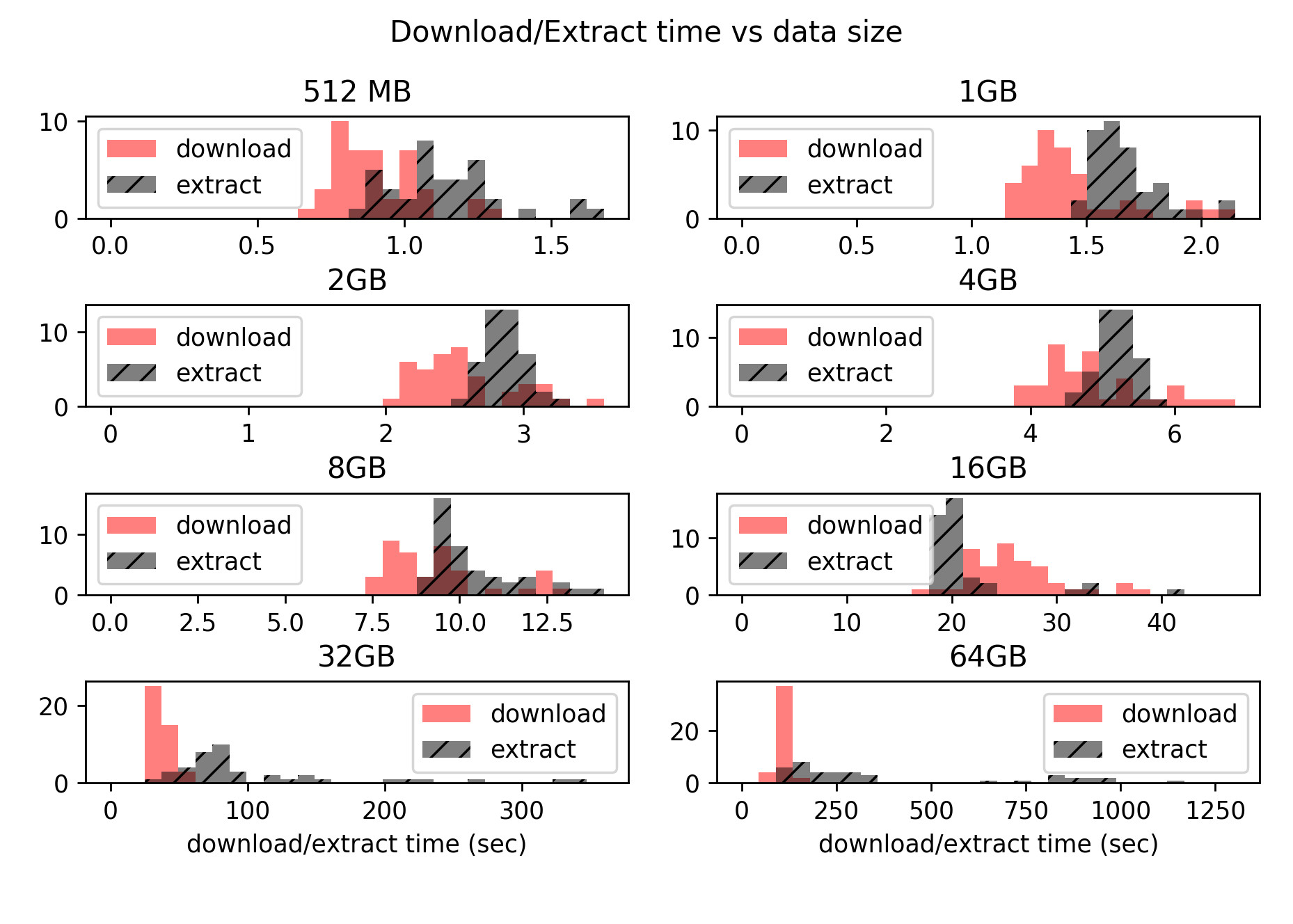}
      \caption{A histogram of the download and extracting times of multiple data sizes on the \texttt{GINA} worker nodes. Download and extract times are comparable for data up to 8GB, however above that, the extracting time dominates.  }
	\label{fig:dl_hist}
\end{figure}

\begin{figure}
    \includegraphics[width=0.95\linewidth]{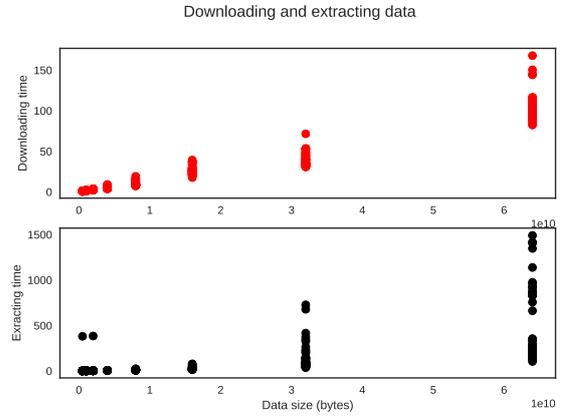}
      \caption{A scatter plot of the download and extracting times of multiple data sizes on the \texttt{GINA} worker nodes. The difference between download and extract time for the 32 and 64 GB data sets can be seen.  }
	\label{fig:dl_plot}
\end{figure}

Figure \ref{fig:dl_hist} shows the histogram of the download tests, and Figure \ref{fig:dl_plot} displays the tests as a function of data size. Both figures show that extracting of the 32 and 64GB data sets has more slow outliers than the downloading of this data. 

We fit a power law model to the time taken to transfer and unpack the data. In this case, we also consider the 75th percentile of these times in order to capture the majority of runs and ignore outliers. The plot of the data and our model can be seen in Figure \ref{fig:dl_plot} and the model is in Equation \ref{eq:download_model}, as a function of the input data size,
$\mathcal{S} $ in gigabytes.

\begin{figure}
    \includegraphics[width=0.95\linewidth]{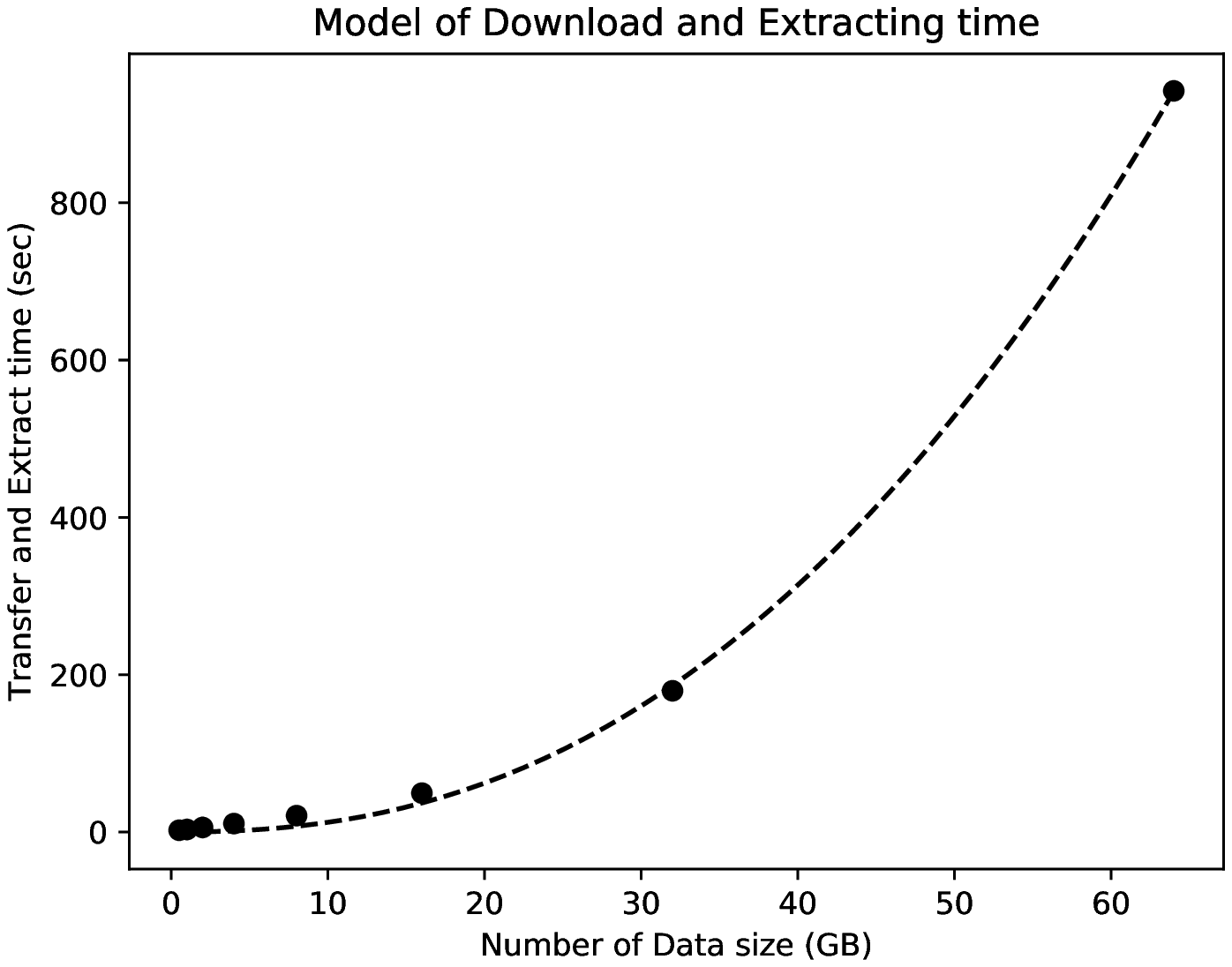}
      \caption{Fit of an exponential model to the Download and Extraction time for different data sizes. For the transfer overhead, we took the 75th percentile from the data shown in Figure \ref{fig:dl_hist}. The model in Equation \ref{eq:download_model} is shown in a dahsed line. }
	\label{fig:dl_ex_model}
\end{figure}

\begin{equ}
\begin{equation}
  T=5.918\times10^{20}\cdot \mathcal{S}^{2.336}
  \label{eq:download_model}
\end{equation}
\caption{Model of the downloading and extracting time as a function of the data size ($\mathcal{S}$) in bytes.}
\end{equ}

\subsection{Comparison with production runs}
Over the past two years, the LOFAR software has been running in production and collecting data on run time for each processing step. We have saved detailed logs for these runs starting in July 2018.  We can compare this to the isolated model in order to determine the overhead incurred by processing LOFAR data on shared nodes. 

\begin{figure}
    \includegraphics[width=0.95\linewidth]{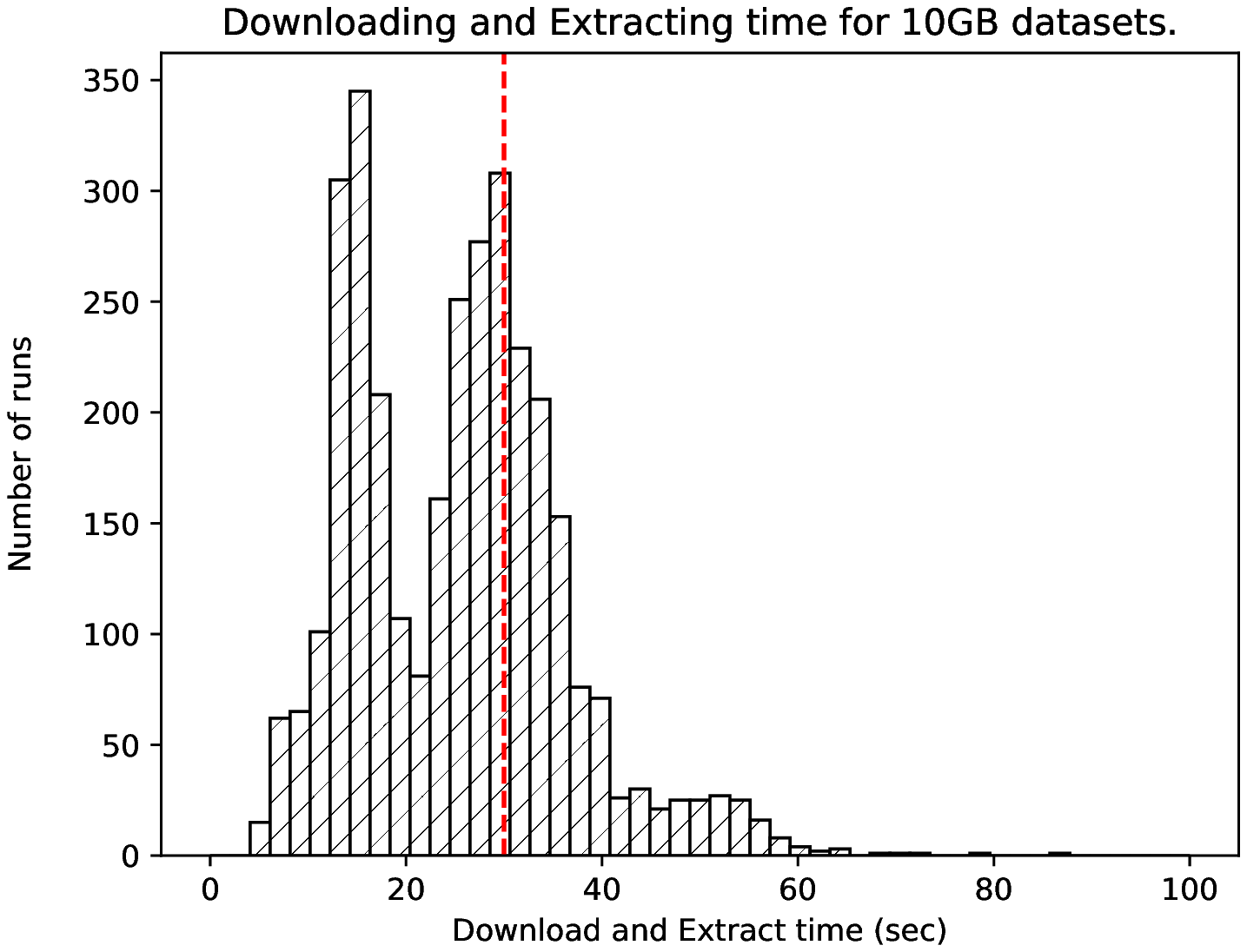}
      \caption{Downloading and extracting time for 10 1GB data sets performed in our production environment. Data from this test ranges from July 2018 to January 2019. The dashed red line shows the prediction obtained from section \ref{sec:results_dl}. We see a bimodal distribution corresponding to 10 GB data (right peak) and data averaged further to 5GB (left peak).}
	\label{fig:prod_dl_10}
\end{figure}

\begin{figure}
    \includegraphics[width=0.95\linewidth]{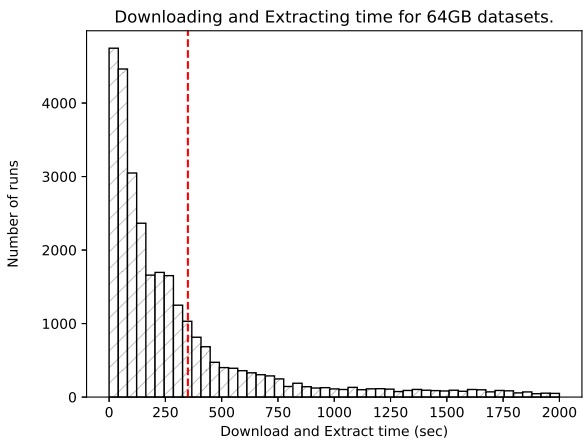}
      \caption{Downloading and extracting time for a 64GB data set performed in our production environment. Data from this test ranges from 07/2018-01/2019. The dashed red line shows the prediction obtained from Figure \ref{fig:gsmcalsolve_size} in Section \ref{sec:results_dl}. }
	\label{fig:prod_dl_64}
\end{figure}

Using the logs recorded by our processing launcher \footnote{GRID\_PiCaS\_Launcher, \raggedright\url{https://github.com/apmechev/GRID\_picastools}}, we made plots showing the processing time for the downloading and extracting, and for the slowest steps, {\fontfamily{qcr}\selectfont ndppp\_prepcal} and {\fontfamily{qcr}\selectfont gsmcal\_solve}. 
The results are shown in Figures \ref{fig:prod_dl_10} and \ref{fig:prod_dl_64}. We include predicted extract times from Section \ref{sec:results_dl} as vertical dashed lines for both plots. The agreement between our model and production runs are an encouraging result for future software performance modelling. 

\begin{figure}
    \includegraphics[width=0.95\linewidth]{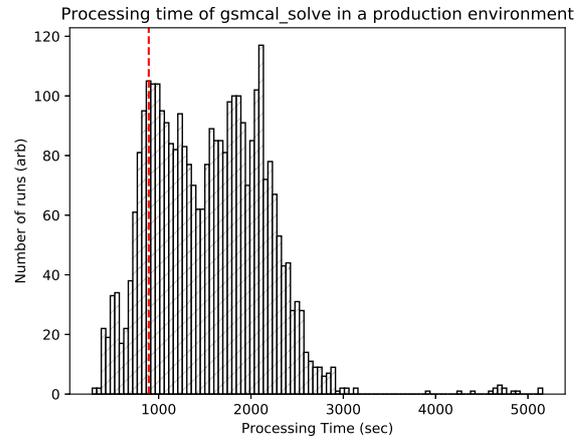}
      \caption{Processing time for the {\fontfamily{qcr}\selectfont gsmcal\_solve} step in a production environment. Data from this test ranges from 07/2018-01/2019. The dashed red line shows the prediction for a 1GB run, obtained from section \ref{sec:results_dl}. We see two distributions, which correspond to data averaged to 1GB and 512 MB.  It should be noted that the left peak corresponds to 512MB data, as seen in Figure \ref{fig:prod_gsmcal_times}.}
	\label{fig:prod_gsmcal}
\end{figure}

\begin{figure}
    \includegraphics[width=0.95\linewidth]{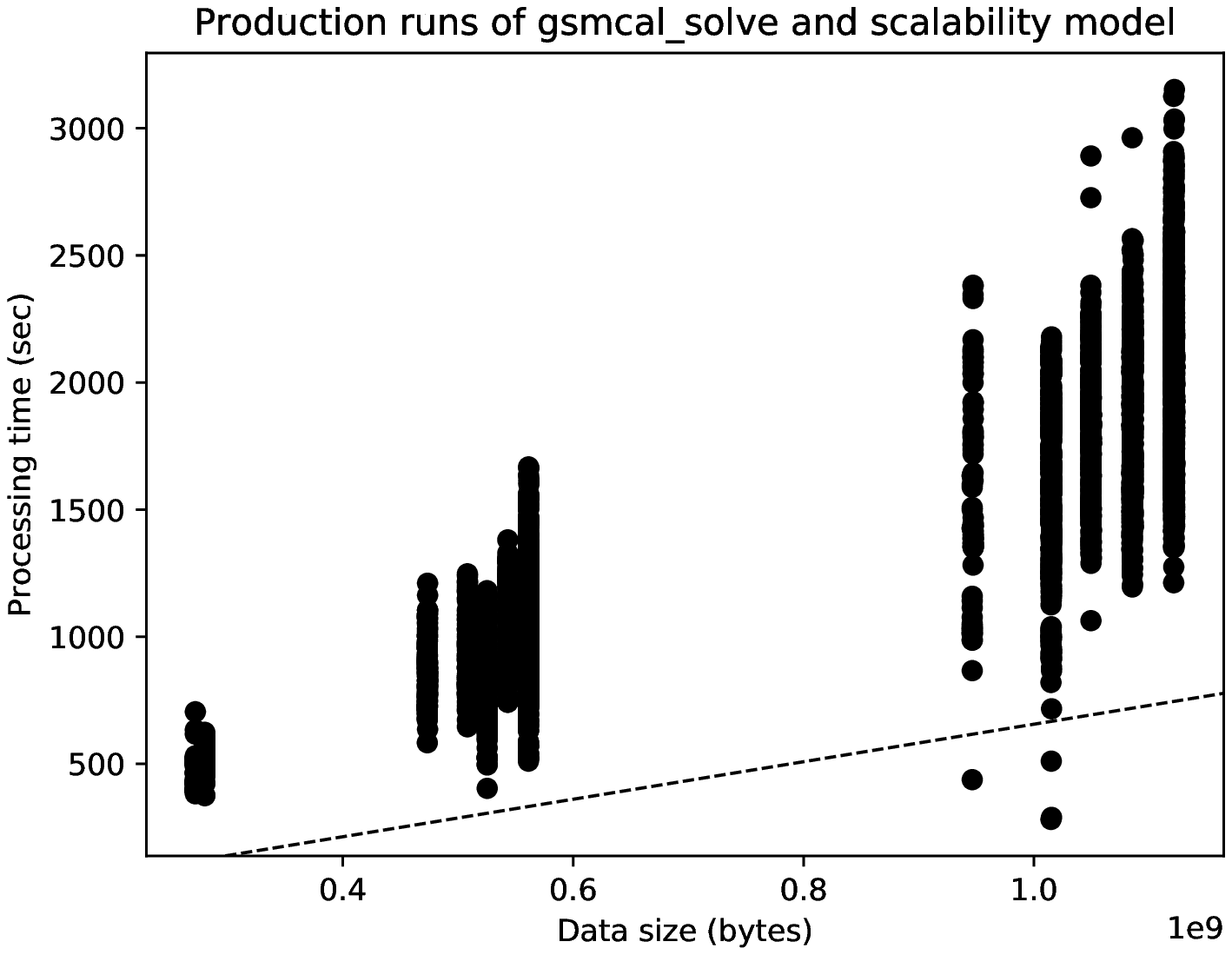}
      \caption{The scalability model for processing data through the {\fontfamily{qcr}\selectfont gsmcal\_solve} step, shown in a dashed line. The scatter plot shows the performance for production runs of this step between July 2018 and January 2019. The two large clusters are for data products that are 1.0 and 2.0 GB respectively. }
	\label{fig:prod_gsmcal_times}
\end{figure}

Finally, we present Figure \ref{fig:prod_gsmcal} which shows a comparison of {\fontfamily{qcr}\selectfont gsmcal\_solve} run times and our model's prediction for a 1GB data set.  Figure \ref{fig:prod_gsmcal_times} plots the processing time vs data size for these production runs and includes the model from Equation \ref{eq:gsmcalsolve}. The significant overhead incurred on a shared infrastructure can be noted. 

\subsection{Complete Scalability Model}
To incorporate all our data into a complete model, we consider the slowdown of each parameter as a multiplier to the time taken to process our base run. We incorporate the models for each parameter above for the model of the run time. We add the transfer and queuing time to the processing time to obtain a final function of all our parameters. We can use this function to predict the processing time for an arbitrary data set. 

The final performance model for the slowest steps, {\fontfamily{qcr}\selectfont gsmcal\_solve}, {\fontfamily{qcr}\selectfont dpppconcat}  and {\fontfamily{qcr}\selectfont predict\_ateam} are in Equation \ref{eq:final_models}. 

\begin{equ*}[!t]
\normalsize
\begin{subequations}

\begin{equation}
 \begin{split}
   t_{infrastructure} & = \begin{cases}
        49.3\cdot\mathcal{N}+ 120 &|\mathcal{N}\leq4\\
        726\cdot\mathcal{N}-3071 & |\mathcal{N}>4
       \end{cases} \\
       & + 2 \cdot 0.056\cdot \mathcal{S}^{2.336} 
       \end{split}
 \end{equation}
     
\begin{equation}
    \begin{split}
   t_{gsmcal\_solve} & = t_{infrastructure} \\
       & + [3566\cdot \frac{1}{3.012}\mathcal{F}^{-0.854} \cdot (0.1412+\frac{0.8589}{\mathcal{N}}) ]  \cdot
          \begin{cases} 
             7.38\cdot10^{-7}\mathcal{S} | \mathcal{S}\leq16\\
             1.04\cdot10^{-6}\mathcal{S} | \mathcal{S}>16 
          \end{cases}\\
       \end{split}
 \label{eq:full_model_gsmcal}
 \end{equation}
 
 \begin{equation}
 \begin{split}
 t_{dpppconcat} &= t_{infrastructure} \\
       & + 3.51\times10^{-8}\mathcal{S}+4.20\times10^1
       \end{split}
       \label{eq:full_model_dppconcat}
 \end{equation}
 
  \begin{equation}
   \begin{split}
    t_{predict\_ateam}  &= t_{infrastructure} \\
       & + 5.19\times10^{-8}\mathcal{S}+4.20\times10^1 
       \end{split}\label{eq:full_model_predict_ateam}
  \end{equation}
\end{subequations}

\caption{Model of the total time of the most computationally expensive steps for the parameters $\mathcal{N}$, Number of CPUs; $\mathcal{S}$, Size of data in bytes and $\mathcal{F}$, cutoff calibration model flux in Jansky. These models include processing times, as well as infrastructure overheads. As the model for the queuing, downloading and uploading time does not change for different processing steps, we decide to keep it separate for clarity. The complete scalability models for the rest of the steps can be derived similarly, however are omitted here as they consist of the minority of processing time for LOFAR DI processing.  }
\label{eq:final_models}
\end{equ*}

\section{Discussions and Conclusions}\label{sec:discussions}

The goal of this work is to understand the performance of the LOFAR Direction Independent Pipeline as processing parameters are changed. We modify several parameters and compare the wall clock time taken to process the data. Finally, we study data from queuing jobs and downloading data, in order to fully model infrastructure overheads. We compare our model with past runs of the software and discuss the results and implications. We discuss the utility of this model for current and upcoming LOFAR projects. Finally, we note the effectiveness of this modelling technique for understanding the performance of large-scale processing of astronomical data.

\subsection{Software Performance}

We performed several tests to determine the scalability of LOFAR processing with respect to several parameters. We outline our findings below as well as their implications for processing in context of the LOFAR surveys and other LOFAR projects. 

\subsubsection{Data Size}
We tested broadband LOFAR data ranging in size by a factor of 64, and discover that all our processing steps scale linearly in time with respect of the input data size. We learn that for input data above 16GB, the slope of our scaling relation is higher than for the smaller data sets. The linear scaling of our processing suggests that projects interested in processing massive LOFAR data sets can scale well in terms of processing time. 

As the calibration step concatenates 10 input subbands, data larger than 16GB shows a higher slope (Figure \ref{fig:gsmcalsolve_size}), meaning they take longer to process than smaller data. We compared the 10GB run and the 64GB run, which showed that the latter fills the system's page cache, while the former does not. It is likely that this memory usage is related to the overhead seen in Figure \ref{fig:prod_gsmcal_times} and 16GB break seen in Figure \ref{fig:gsmcalsolve_size}. Including this break in the model also helps make a more accurate processing time prediction as fitting a single linear model would have a large negative y-intercept, predicting negative processing times for data smaller than 2GB. 

Our analysis shows the following results. Overall, the slowest step was the {\fontfamily{qcr}\selectfont gsmcal\_solve} step, and its run time scales more strongly with data size than the other steps (equation \ref{eq:gsmcalsolve} has the steepest slope). This suggests that as data sizes increase, {\fontfamily{qcr}\selectfont gsmcal\_solve} will increasingly dominate the processing time over the other steps (As seen in Figures \ref{fig:predict_ateam} and \ref{fig:gsmcalsolve_size}). This effect is especially prominent for data larger than 16GB (160GB when 10 subbands are concatenated). As such, it is recommended to avoid calibration of data larger than 160GB. This limitation suggests that science requiring operations on non-averaged data sets, such as long-baseline imaging, will require significant computational requirements for high-fidelity images.

\subsubsection{Calibration Model Size}
We discover that the calibration time scales linearly in as a function of the length of the calibration model, however as a power law with respect to the model's cutoff sensitivity. This is because of the (expected) power law relation between the number of sources and cutoff sensitivity, seen in Figure \ref{fig:skymodel_size}. We can use this discovery to accelerate the processing time by increasing the flux cutoff to the LOFAR direction independent calibration to 0.5 Jy. Doing so will execute the calibration step in 60\% of the time, saving 83 CPU-h per run. Over the remaining 2000 \texttt{prefactor} runs left in the LOTSS project, this change in sensitivity will save more than 167k CPU-hours. 

Figures \ref{fig:skymodel_images} show a data set calibrated with sky models with cutoff sensitivities listed in Table \ref{table:skymodel_RMS}, and figures  \ref{fig:skymodel_rcalib_004_03} and \ref{fig:skymodel_rcalib_08_15} show the calibration solutions obtained by calibrating with skymodels of cutoff ranging from 0.05 Jy to 1.5 Jy. These results suggest that performing gain calibration with less complex, and thus smaller, calibration models will not degrade image and solution quality while taking less than 20\% of processing time. Table \ref{table:skymodel_RMS} also confirms this result. 

While this result is encouraging, there are caveats suggesting future study is required. The results we present are for a single observation, and has not been processed through the Direction Dependent Calibration pipeline. This pipeline produces final images used for scientific research. Future work will need to confirm that smaller calibration models used in the \texttt{prefactor} pipeline do not degrade the quality of these final images. Nevertheless if this result holds, the calibration model threshold can be decreased for a wide range of LOFAR projects, significantly saving processing time and computing resources. 

\subsubsection{Comparison with production runs}

When comparing our model's prediction with real processing runs over the past six months, we note that there are considerable overheads when running on a shared infrastructure vs. when processing data on an isolated (Figure \ref{fig:prod_gsmcal_times}). The overhead in processing is roughly a factor of two-three from our model. This discrepancy suggests that a model for {\fontfamily{qcr}\selectfont gsmcal\_solve} needs to be built using data when running on a shared environment, to better predict processing time. 

\subsection{Infrastructure Performance}
We tested downloading and extracting LOFAR data of various sizes. Both downloading and extracting are shown to be linear in time with respect to the data size for data up to 32 GB. Beyond those sizes, there is more scatter in data extraction due to high file-system load. This is because load on the worker node's file-system can be unpredictable and can affect the data extraction times negatively. Nevertheless, when comparing our extraction tests and processing for the past 6 months, the predictions by our models (Figure \ref{fig:dl_hist}) correspond to the production runs (Figures \ref{fig:prod_dl_10} and \ref{fig:prod_dl_64}).

Part of the LOFAR SKSP processing is done on shared infrastructure, which requires requesting processing time ahead of time for each grant period. Being able to predict the amount of resources required to process data each grant period is required to make a reliable estimate on what resources to request. These results can be used by other projects sharing the SURFsara \texttt{GINA} cluster to predict their processing time before submitting jobs. 

\section{Applications and Conclusions}

Our performance model shows that it is possible to predict the processing time and computational resources used by a complex astronomical pipeline. Our results suggest that LOFAR LoTSS processing can be further optimized without sacrificing quality of the final product. Additionally, our results are transferable to other scientific pipelines that process LOFAR data with different parameters.  Any pipeline that performs gain calibration or application of calibration solutions will benefit from these results. 

In order to provide LOFAR processing as a service to scientific users, we need to estimate the processing time for each request. We need this estimation in order to determine whether the user has sufficient resources left in their resource pool. Knowing the performance of the software pipelines as a function of the input parameters will help predict the run time for each request and the resources consumed. Knowing this will make it possible to notify users how long the request will take and how much of their quota will be depleted. It is important to note that while this model is specific to LOFAR processing, the method we detail can be used by other scientific teams in order to predict the computational requirements for their pipelines. Doing this is necessary if large scale scientific processing is to be offered as a service to the wider community.

Finally, a performance model of the LOFAR software will help make predictions on the time and resources needed to process data for other telescopes such as the Square Kilometer Array (SKA). Once operational, the SKA is expected to produce Exabytes of data per year. Processing this data efficiently requires understanding the scalability of the software performance to facilitate scheduling and resource allotment. Overall, we show that our method helps guide algorithm development in radio astronomy, can be used to predict resource usage by complex pipelines and will be promising in optimizing data processing by future telescopes.

\appendix

\renewcommand{\theequ}{\Alph{section}.\arabic{equ}}

\section{Calibration Solutions for the sky model tests }\label{ap:calib_solutions}
\begin{figure}
    \includegraphics[width=0.95\linewidth]{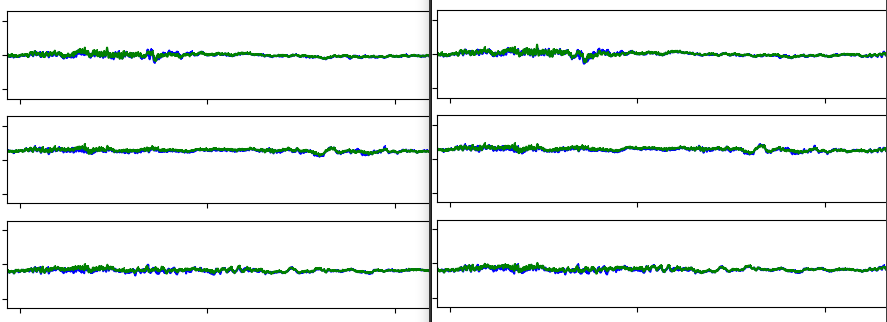}
      \caption{The calibration (phase) solutions for the test dataset obtained when calibrating with sky models of 0.05 Jy cutoff (left) and 0.3Jy cutoff (right). The data shows the phase solutions for baselines including stations CS003HBA0, CS003HBA1 and CS004HBA0, with respect to the reference station, CS001HBA0. The right solutions were obtained using the production calibration model. We do not see any improvement in results in the left figure, which took twice as long to obtain.}
	\label{fig:skymodel_rcalib_004_03}
\end{figure}

\begin{figure}
    \includegraphics[width=0.95\linewidth]{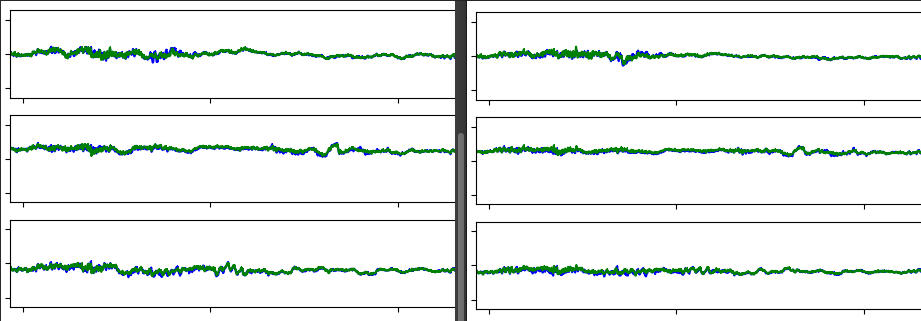}
      \caption{The calibration (phase) solutions for the test dataset obtained when calibrating with sky models of 0.8 Jy cutoff (left) and 1.5Jy cutoff (right). The data shows the phase solutions for baselines including stations CS003HBA0, CS003HBA1 and CS004HBA0, with respect to the reference station, CS001HBA0. We can see that the calibration solutions shown here are not significantly different than those shown in \ref{fig:skymodel_rcalib_004_03}, despite taking a fraction of the processing time.  }
	\label{fig:skymodel_rcalib_08_15}
\end{figure}

The output of the calibration step is a data set corrected for direction independent effects, as well as a set of calibration solutions. Figures \ref{fig:skymodel_rcalib_004_03} and \ref{fig:skymodel_rcalib_08_15} show the calibration solutions for core stations obtained when calibrating with sky models with minimum flux cutoffs of 0.05, 0.3, 0.8 and 1.5Jy. Much like in Figure \ref{fig:skymodel_images}, we can see that there is no significant difference between the calibration solutions for these stations. As a note, the naming scheme for LOFAR stations is CS/RS for core/remote stations, three digits for station number, HBA0/HBA1 for High Band antennas and LBA0/LBA1 for low-band antennas. The 0,1 suffixes correspond to sub-arrays in the core stations which can be correlated separately. Additionally, the CS/RS is replaced with the 2-letter country code for international stations\citep{staiton_data_cookbook}.

\begin{figure}
    \includegraphics[width=0.9\linewidth]{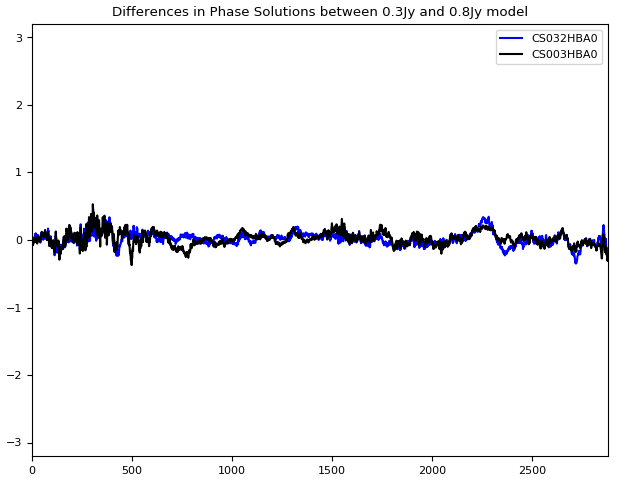}
      \caption{Difference of phase solutions between calibrations with the 0.3Jy and 0.8Jy sky models. The solutions for both stations are around zero phase for the duration of the observation.}
	\label{fig:diffs_solutions}
\end{figure}

We compare the phase solutions for stations CS032HBA0 and CS003HBA0 and the reference station for two different calibrations in Figure \ref{fig:diffs_solutions}. We note that this difference is within 0.3 rad for the entire observation. Combined with the results in the other two plots, our results suggest that the calibration solutions do not degrade when calibration is done with a smaller sky model.

\section{Parametric model parameters and fit accuracy}\label{ap:model_params}

In this section, we note the uncertainties to the models fit in Equations \ref{eq:runtime_size_models}-\ref{eq:download_model}. 

%Resetting the equation environment to match the equ number
\numberwithin{equation}{section}
\setcounter{equation}{6}
\renewcommand{\theequation}{\Alph{section}.\arabic{equation}}

\subsection{Fits quality of run time vs input size model}

The models of the processing time vs input size were fit as a linear regression. In this work we present such models for the {\fontfamily{qcr}\selectfont  gsmcal\_solve}, {\fontfamily{qcr}\selectfont gsmcal\_apply}, {\fontfamily{qcr}\selectfont dpppconcat}, {\fontfamily{qcr}\selectfont predict\_ateam} and {\fontfamily{qcr}\selectfont ateamcliptar}, the five slowest steps. The resulting models, calculated by the \texttt{scipy linregress}\citep{scipy} routine, are shown in Equation \ref{eq:runtime_size_models}. We present the $R^2$ values, P values and standard error below, in Table \ref{table:fits_size}.

\begin{table}[ht!]
\centering
\begin{tabular}{||p{2.2cm}| c | c|p{2cm}||} 
 \hline
 \texttt{prefactor} step & $R^2$ & P value & Standard Error \\ %[0.5ex]
 \hline
 predict\_ateam & 0.996   & 0                    & $1.92\times10^{-10}$    \\ 
 \hline
 ateamcliptar   & 0.979   & 0                    & $3.94\times10^{-11}$    \\ 
 \hline
 dpppconcat     & 0.999   & $1.2\times10^{-128}$ & $1.78\times10^{-10}$    \\ 
 \hline
 gsmcal\_solve $<=$16GB  & 0.995   & $3.12\times10^{-75}$ & $6.80\times10^{-9}$     \\ 
 \hline
 gsmcal\_solve $>$16GB  & 0.951   & $7.07\times10^{-40}$ & $1.58\times10^{-8}$     \\ 
 \hline
 gsmcal\_apply  & 0.989   & $5.6\times10^{-82}$  & $3.12\times10^{-10}$    \\ 

\hline
\end{tabular}
\caption{Fit parameters for the models in Equation \ref{eq:runtime_size_models}. }
\label{table:fits_size}
\end{table}

\subsection{Fit of run time vs calibration model flux cutoff }

The run time vs Flux cutoff model shown in Equation \ref{eq:skymodel_flux} is defined by the equation $y=a\cdot x^{-k}$ and two parameters, $a$ and $k$. The covariance matrix for these two parameters is shown in Equation \ref{eq:cov_Flux}. The standard deviation for the fit of the parameters $a$ and $k$ is 26.134 and $7.624\times10^{-3}$ respectively.

\begin{equ}
\begin{equation}
  \begin{bmatrix}
    6.83\times10^{2}  &  -1.94\times10^{-1} \\
   -1.94\times10^{-1} &   5.81\times10^{-5} \\
\end{bmatrix}
\end{equation}
\caption{The covariance matrix of the parameters in model in Equation \ref{eq:skymodel_flux}.}
\label{eq:cov_Flux}
\end{equ}

\subsection{Fit of the NCPU model }
The covariance matrix for the fit parameters of equation \ref{eq:gsmcal_NCPU}, $a$ and $k$ in $y=a+\frac{k}{\mathcal{N}}$ are shown in Equation \ref{eq:cov_NCPU}. The standard deviation of the fits for $a$ and $k$ are 13.11 and 48.20 respectively. 

\begin{equ}
\begin{equation}
  \begin{bmatrix}
    171.94 & -504.11 \\
    -504.11 & 2322.95
\end{bmatrix}
\end{equation}
\caption{The covariance matrix for the parameters for the model predicting run time vs Number of CPUs used, shown in Equation \ref{eq:gsmcal_NCPU}.}
\label{eq:cov_NCPU}
\end{equ}

\subsection{Fit for the queuing time model}

The statistics of the model fit parameters for the queuing time model (Equation \ref{eq:queue_model}) are in Table \ref{table:fits_queue}. 
The queuing model is fit to the $75^{th}$ percentile of the queuing times for each parameter step. Since this results in a single number for each step, the model's P values are larger than the models from the other sections. 

\begin{table}[ht!]
\centering
\begin{tabular}{||p{2.2cm}| c | c|p{2cm}||} 
 \hline
 Value of $\mathcal{N}$ & $R^2$ & P value & Standard Error \\ %[0.5ex]
 \hline
 $\mathcal{N}\leq4$ & 0.382   & 0.381                   & 37.086    \\ 
 $\mathcal{N}>4$    & 0.986   & 0.075                   & 86.293    \\ 
\hline
\end{tabular}
\caption{Goodness of fit parameters for the model in Equation \ref{eq:queue_model}. Since the model is split into two parts, we treat each section as a single linear model.}
\label{table:fits_queue}
\end{table}

\subsection{Fit of the download and extract model }
Equation \ref{eq:cov_downl} shows the covariance matrix for the two parameters  $a$ and $k$, $y=a\times10^{k}$ with the best fit values shown in Equation \ref{eq:download_model}. The standard deviations of the fits for $a$ and $k$ are 0.016 and 0.068 respectively. 

\begin{equ}
\begin{equation}
  \begin{bmatrix}
    2.53\times10^{-4} & 1.08\times10^{-3} \\
    1.08\times10^{-3} & 4.60\times10^{-3}
\end{bmatrix}
\end{equation}
\caption{The covariance matrix for the parameters for the model for Download and Extract time, shown in Equation \ref{eq:download_model}.}
\label{eq:cov_downl}
\end{equ}

\section*{Acknowledgments}
APM would like to acknowledge the support from the NWO/DOME/IBM programme ``Big Bang Big Data: Innovating ICT as a Driver For Astronomy'', project \#628.002.001.

This work was carried out on the Dutch national e-infrastructure with the support of SURF
Cooperative through grants e-infra 160152 and e-infra 180169.

\section*{References}
\bibliography{bibliography}

\begin{thebibliography}{30}
\providecommand{\natexlab}[1]{#1}
\providecommand{\url}[1]{\texttt{#1}}
\expandafter\ifx\csname urlstyle\endcsname\relax
  \providecommand{\doi}[1]{doi: #1}\else
  \providecommand{\doi}{doi: \begingroup \urlstyle{rm}\Url}\fi

\bibitem[Aguado~Sanchez et~al.(2008)Aguado~Sanchez, Bloomer, Buncic, Franco,
  Klemer, and Mato]{cvmfs2008}
C.~Aguado~Sanchez, J.~Bloomer, P.~Buncic, L.~Franco, S.~Klemer, and P.~Mato.
\newblock {CVMFS} a file system for the {CernVM} virtual appliance.
\newblock In \emph{Proceedings of XII Advanced Computing and Analysis
  Techniques in Physics Research}, volume~1, page~52, 2008.

\bibitem[Barnes et~al.(2008)Barnes, Rountree, Lowenthal, Reeves, De~Supinski,
  and Schulz]{barnes2008regression}
B.~J. Barnes, B.~Rountree, D.~K. Lowenthal, J.~Reeves, B.~De~Supinski, and
  M.~Schulz.
\newblock A regression-based approach to scalability prediction.
\newblock In \emph{Proceedings of the 22nd annual international conference on
  Supercomputing}, pages 368--377. ACM, 2008.

\bibitem[Calotoiu et~al.(2013)Calotoiu, Hoefler, Poke, and
  Wolf]{scalability_bugs}
A.~Calotoiu, T.~Hoefler, M.~Poke, and F.~Wolf.
\newblock Using automated performance modeling to find scalability bugs in
  complex codes.
\newblock In \emph{Proceedings of the International Conference on High
  Performance Computing, Networking, Storage and Analysis}, page~45. ACM, 2013.

\bibitem[Carrington et~al.(2006)Carrington, Snavely, and
  Wolter]{synthetic_memory_prediction}
L.~Carrington, A.~Snavely, and N.~Wolter.
\newblock A performance prediction framework for scientific applications.
\newblock \emph{Future Generation Computer Systems}, 22\penalty0 (3):\penalty0
  336--346, 2006.

\bibitem[Castiglione et~al.(2014)Castiglione, Gribaudo, Iacono, and
  Palmieri]{mean_field_modeling}
A.~Castiglione, M.~Gribaudo, M.~Iacono, and F.~Palmieri.
\newblock Exploiting mean field analysis to model performances of big data
  architectures.
\newblock \emph{Future Generation Computer Systems}, 37:\penalty0 203 -- 211,
  2014.
\newblock ISSN 0167-739X.
\newblock \doi{https://doi.org/10.1016/j.future.2013.07.016}.
\newblock URL
  \url{http://www.sciencedirect.com/science/article/pii/S0167739X13001611}.
\newblock Special Section: Innovative Methods and Algorithms for Advanced
  Data-Intensive Computing Special Section: Semantics, Intelligent processing
  and services for big data Special Section: Advances in Data-Intensive
  Modelling and Simulation Special Section: Hybrid Intelligence for Growing
  Internet and its Applications.

\bibitem[Dijkema(2017)]{cookbook}
T.~J. Dijkema.
\newblock {LOFAR} {I}maging {C}ookbook.
\newblock Available at
  \url{http://www.astron.nl/sites/astron.nl/files/cms/lofar_imaging_cookbook_v19.pdf
  }, 2017.

\bibitem[Horneffer et~al.(2018)Horneffer, Williams, Shimwell, Roskowinski,
  Rafferty, Mechev, Dziełak, Bourke, Dijkema, Hardcastle, and
  Sabater]{prefactor_zenodo}
A.~Horneffer, W.~Williams, T.~Shimwell, C.~Roskowinski, D.~Rafferty, A.~Mechev,
  M.~Dziełak, S.~Bourke, T.~J. Dijkema, M.~Hardcastle, and J.~Sabater.
\newblock apmechev/prefactor: {LOTSS} {D}ata {Release} 1, Nov. 2018.
\newblock URL \url{https://doi.org/10.5281/zenodo.1487962}.

\bibitem[Intema et~al.(2017)Intema, Jagannathan, Mooley, and Frail]{tgssadr}
H.~Intema, P.~Jagannathan, K.~Mooley, and D.~Frail.
\newblock The {GMRT} 150 {MH}z all-sky radio survey-first alternative data
  release {TGSS ADR1}.
\newblock \emph{Astronomy \& Astrophysics}, 598:\penalty0 A78, 2017.

\bibitem[Jones et~al.(2001--)Jones, Oliphant, Peterson, et~al.]{scipy}
E.~Jones, T.~Oliphant, P.~Peterson, et~al.
\newblock {SciPy}: Open source scientific tools for {Python}, 2001--.
\newblock URL \url{http://www.scipy.org/}.
\newblock [Online; accessed \today].

\bibitem[Kavulya et~al.(2010)Kavulya, Tan, Gandhi, and
  Narasimhan]{mapreduce_analysis}
S.~Kavulya, J.~Tan, R.~Gandhi, and P.~Narasimhan.
\newblock An analysis of traces from a production mapreduce cluster.
\newblock In \emph{2010 10th IEEE/ACM International Conference on Cluster,
  Cloud and Grid Computing}, pages 94--103, May 2010.
\newblock \doi{10.1109/CCGRID.2010.112}.

\bibitem[Kazemi et~al.(2011)Kazemi, Yatawatta, Zaroubi, Lampropoulos, De~Bruyn,
  Koopmans, and Noordam]{radio_visibility_sage}
S.~Kazemi, S.~Yatawatta, S.~Zaroubi, P.~Lampropoulos, A.~De~Bruyn, L.~Koopmans,
  and J.~Noordam.
\newblock Radio interferometric calibration using the sage algorithm.
\newblock \emph{Monthly Notices of the Royal Astronomical Society},
  414\penalty0 (2):\penalty0 1656--1666, 2011.

\bibitem[Kuperberg et~al.(2008)Kuperberg, Krogmann, and
  Reussner]{performance_prediction}
M.~Kuperberg, K.~Krogmann, and R.~Reussner.
\newblock Performance prediction for black-box components using reengineered
  parametric behaviour models.
\newblock In \emph{International Symposium on Component-Based Software
  Engineering}, pages 48--63. Springer, 2008.

\bibitem[Levanda and Leshem(2010)]{app_synth}
R.~Levanda and A.~Leshem.
\newblock Synthetic aperture radio telescopes.
\newblock \emph{Signal Processing Magazine, IEEE}, 27:\penalty0 14–29, 02
  2010.
\newblock \doi{10.1109/MSP.2009.934719}.

\bibitem[Marco et~al.(2010)Marco, Fabio, Alvise, Antonia, Alessio, Francesco,
  Alessandro, Elisabetta, Salvatore, and Luca]{glite-wms}
C.~Marco, C.~Fabio, D.~Alvise, G.~Antonia, G.~Alessio, G.~Francesco,
  M.~Alessandro, M.~Elisabetta, M.~Salvatore, and P.~Luca.
\newblock The glite workload management system.
\newblock In \emph{Journal of Physics: Conference Series}, volume 219, page
  062039. IOP Publishing, 2010.

\bibitem[{Mechev} et~al.(2017){Mechev}, {Oonk}, {Danezi}, {Shimwell},
  {Schrijvers}, {Intema}, {Plaat}, and {Rottgering}]{mechev17}
A.~{Mechev}, J.~B.~R. {Oonk}, A.~{Danezi}, T.~W. {Shimwell}, C.~{Schrijvers},
  H.~{Intema}, A.~{Plaat}, and H.~J.~A. {Rottgering}.
\newblock {An {A}utomated {S}calable {F}ramework for {D}istributing {R}adio
  {A}stronomy {P}rocessing {A}cross {C}lusters and {C}louds}.
\newblock In \emph{Proceedings of the International Symposium on Grids and
  Clouds (ISGC) 2017, held 5-10 March, 2017 at Academia Sinica, Taipei, Taiwan
  (ISGC2017). Online at
  \url{https://pos.sissa.it/cgi-bin/reader/conf.cgi?confid=293}, id.2}, page~2,
  Mar. 2017.

\bibitem[{Offringa} et~al.(2014){Offringa}, {McKinley}, {Hurley-Walker},
  {Briggs}, {Wayth}, {Kaplan}, {Bell}, {Feng}, {Neben}, {Hughes}, {Rhee},
  {Murphy}, {Bhat}, {Bernardi}, {Bowman}, {Cappallo}, {Corey}, {Deshpande},
  {Emrich}, {Ewall-Wice}, {Gaensler}, {Goeke}, {Greenhill}, {Hazelton},
  {Hindson}, {Johnston-Hollitt}, {Jacobs}, {Kasper}, {Kratzenberg}, {Lenc},
  {Lonsdale}, {Lynch}, {McWhirter}, {Mitchell}, {Morales}, {Morgan},
  {Kudryavtseva}, {Oberoi}, {Ord}, {Pindor}, {Procopio}, {Prabu}, {Riding},
  {Roshi}, {Shankar}, {Srivani}, {Subrahmanyan}, {Tingay}, {Waterson},
  {Webster}, {Whitney}, {Williams}, and {Williams}]{wsclean}
A.~R. {Offringa}, B.~{McKinley}, N.~{Hurley-Walker}, F.~H. {Briggs}, R.~B.
  {Wayth}, D.~L. {Kaplan}, M.~E. {Bell}, L.~{Feng}, A.~R. {Neben}, J.~D.
  {Hughes}, J.~{Rhee}, T.~{Murphy}, N.~D.~R. {Bhat}, G.~{Bernardi}, J.~D.
  {Bowman}, R.~J. {Cappallo}, B.~E. {Corey}, A.~A. {Deshpande}, D.~{Emrich},
  A.~{Ewall-Wice}, B.~M. {Gaensler}, R.~{Goeke}, L.~J. {Greenhill}, B.~J.
  {Hazelton}, L.~{Hindson}, M.~{Johnston-Hollitt}, D.~C. {Jacobs}, J.~C.
  {Kasper}, E.~{Kratzenberg}, E.~{Lenc}, C.~J. {Lonsdale}, M.~J. {Lynch}, S.~R.
  {McWhirter}, D.~A. {Mitchell}, M.~F. {Morales}, E.~{Morgan},
  N.~{Kudryavtseva}, D.~{Oberoi}, S.~M. {Ord}, B.~{Pindor}, P.~{Procopio},
  T.~{Prabu}, J.~{Riding}, D.~A. {Roshi}, N.~U. {Shankar}, K.~S. {Srivani},
  R.~{Subrahmanyan}, S.~J. {Tingay}, M.~{Waterson}, R.~L. {Webster}, A.~R.
  {Whitney}, A.~{Williams}, and C.~L. {Williams}.
\newblock {WSCLEAN: an implementation of a fast, generic wide-field imager for
  radio astronomy}.
\newblock \emph{MNRAS}, 444:\penalty0 606--619, Oct. 2014.
\newblock \doi{10.1093/mnras/stu1368}.

\bibitem[Sanjay and Vadhiyar(2008)]{grid_perform_model}
H.~Sanjay and S.~Vadhiyar.
\newblock Performance modeling of parallel applications for grid scheduling.
\newblock \emph{Journal of Parallel and Distributed Computing}, 68\penalty0
  (8):\penalty0 1135 -- 1145, 2008.
\newblock ISSN 0743-7315.
\newblock \doi{https://doi.org/10.1016/j.jpdc.2008.02.006}.
\newblock URL
  \url{http://www.sciencedirect.com/science/article/pii/S0743731508000464}.

\bibitem[Shimwell et~al.(2017)Shimwell, R{\"o}ttgering, Best, Williams,
  Dijkema, De~Gasperin, Hardcastle, Heald, Hoang, Horneffer, et~al.]{lotss}
T.~Shimwell, H.~R{\"o}ttgering, P.~N. Best, W.~Williams, T.~Dijkema,
  F.~De~Gasperin, M.~Hardcastle, G.~Heald, D.~Hoang, A.~Horneffer, et~al.
\newblock The {L}{O}{F}{A}{R} {T}wo-metre {S}ky {S}urvey-{I}. {S}urvey
  description and preliminary data release.
\newblock \emph{Astronomy \& Astrophysics}, 598:\penalty0 A104, 2017.

\bibitem[{Shimwell} et~al.(2018){Shimwell}, {Tasse}, {Hardcastle}, {Mechev},
  {Williams}, {Best}, {R{\"o}ttgering}, {Callingham}, {Dijkema}, {de Gasperin},
  {Hoang}, {Hugo}, {Mirmont}, {Oonk}, {Prandoni}, {Rafferty}, {Sabater},
  {Smirnov}, {van Weeren}, {White}, {Atemkeng}, {Bester}, {Bonnassieux},
  {Br{\"u}ggen}, {Brunetti}, {Chy{\.z}y}, {Cochrane}, {Conway}, {Croston},
  {Danezi}, {Duncan}, {Haverkorn}, {Heald}, {Iacobelli}, {Intema}, {Jackson},
  {Jamrozy}, {Jarvis}, {Lakhoo}, {Mevius}, {Miley}, {Morabito}, {Morganti},
  {Nisbet}, {Orr{\'u}}, {Perkins}, {Pizzo}, {Schrijvers}, {Smith}, {Vermeulen},
  {Wise}, {Alegre}, {Bacon}, {van Bemmel}, {Beswick}, {Bonafede}, {Botteon},
  {Bourke}, {Brienza}, {Calistro Rivera}, {Cassano}, {Clarke}, {Conselice},
  {Dettmar}, {Drabent}, {Dumba}, {Emig}, {En{\ss}lin}, {Ferrari}, {Garrett},
  {G{\'e}nova-Santos}, {Goyal}, {G{\"u}rkan}, {Hale}, {Harwood}, {Heesen},
  {Hoeft}, {Horellou}, {Jackson}, {Kokotanekov}, {Kondapally}, {Kunert-
  Bajraszewska}, {Mahatma}, {Mahony}, {Mandal}, {McKean}, {Merloni}, {Mingo},
  {Miskolczi}, {Mooney}, {Nikiel- Wroczy{\'n}ski}, {O'Sullivan}, {Quinn},
  {Reich}, {Roskowi{\'n}ski}, {Rowlinson}, {Savini}, {Saxena}, {Schwarz},
  {Shulevski}, {Sridhar}, {Stacey}, {Urquhart}, {van der Wiel}, {Varenius},
  {Webster}, and {Wilber}]{LOTSS_DR2}
T.~W. {Shimwell}, C.~{Tasse}, M.~J. {Hardcastle}, A.~P. {Mechev}, W.~L.
  {Williams}, P.~N. {Best}, H.~J.~A. {R{\"o}ttgering}, J.~R. {Callingham},
  T.~J. {Dijkema}, F.~{de Gasperin}, D.~N. {Hoang}, B.~{Hugo}, M.~{Mirmont},
  J.~B.~R. {Oonk}, I.~{Prandoni}, D.~{Rafferty}, J.~{Sabater}, O.~{Smirnov},
  R.~J. {van Weeren}, G.~J. {White}, M.~{Atemkeng}, L.~{Bester},
  E.~{Bonnassieux}, M.~{Br{\"u}ggen}, G.~{Brunetti}, K.~T. {Chy{\.z}y},
  R.~{Cochrane}, J.~E. {Conway}, J.~H. {Croston}, A.~{Danezi}, K.~{Duncan},
  M.~{Haverkorn}, G.~H. {Heald}, M.~{Iacobelli}, H.~T. {Intema}, N.~{Jackson},
  M.~{Jamrozy}, M.~J. {Jarvis}, R.~{Lakhoo}, M.~{Mevius}, G.~K. {Miley},
  L.~{Morabito}, R.~{Morganti}, D.~{Nisbet}, E.~{Orr{\'u}}, S.~{Perkins}, R.~F.
  {Pizzo}, C.~{Schrijvers}, D.~J.~B. {Smith}, R.~{Vermeulen}, M.~W. {Wise},
  L.~{Alegre}, D.~J. {Bacon}, I.~M. {van Bemmel}, R.~J. {Beswick},
  A.~{Bonafede}, A.~{Botteon}, S.~{Bourke}, M.~{Brienza}, G.~{Calistro Rivera},
  R.~{Cassano}, A.~O. {Clarke}, C.~J. {Conselice}, R.~J. {Dettmar},
  A.~{Drabent}, C.~{Dumba}, K.~L. {Emig}, T.~A. {En{\ss}lin}, C.~{Ferrari},
  M.~A. {Garrett}, R.~T. {G{\'e}nova-Santos}, A.~{Goyal}, G.~{G{\"u}rkan},
  C.~{Hale}, J.~J. {Harwood}, V.~{Heesen}, M.~{Hoeft}, C.~{Horellou},
  C.~{Jackson}, G.~{Kokotanekov}, R.~{Kondapally}, M.~{Kunert- Bajraszewska},
  V.~{Mahatma}, E.~K. {Mahony}, S.~{Mandal}, J.~P. {McKean}, A.~{Merloni},
  B.~{Mingo}, A.~{Miskolczi}, S.~{Mooney}, B.~{Nikiel- Wroczy{\'n}ski}, S.~P.
  {O'Sullivan}, J.~{Quinn}, W.~{Reich}, C.~{Roskowi{\'n}ski}, A.~{Rowlinson},
  F.~{Savini}, A.~{Saxena}, D.~J. {Schwarz}, A.~{Shulevski}, S.~S. {Sridhar},
  H.~R. {Stacey}, S.~{Urquhart}, M.~H.~D. {van der Wiel}, E.~{Varenius},
  B.~{Webster}, and A.~{Wilber}.
\newblock {The LOFAR Two-metre Sky Survey - II. First data release}.
\newblock \emph{arXiv e-prints}, art. arXiv:1811.07926, Nov. 2018.

\bibitem[Smirnov and Tasse(2015)]{tassesmirnov}
O.~Smirnov and C.~Tasse.
\newblock Radio interferometric gain calibration as a complex optimization
  problem.
\newblock \emph{Monthly Notices of the Royal Astronomical Society},
  449\penalty0 (3):\penalty0 2668--2684, 2015.

\bibitem[Tasse et~al.(2018)Tasse, Hugo, Mirmont, Smirnov, Atemkeng, Bester,
  Hardcastle, Lakhoo, Perkins, and Shimwell]{tasse2018faceting}
C.~Tasse, B.~Hugo, M.~Mirmont, O.~Smirnov, M.~Atemkeng, L.~Bester,
  M.~Hardcastle, R.~Lakhoo, S.~Perkins, and T.~Shimwell.
\newblock Faceting for direction-dependent spectral deconvolution.
\newblock \emph{Astronomy \& Astrophysics}, 611:\penalty0 A87, 2018.

\bibitem[Templon and Bot(2016)]{dutch_einfra}
J.~Templon and J.~Bot.
\newblock The dutch national e-infrastructure.
\newblock {To appear in Proceedings of Science edition of the International
  Symposium on Grids and Clouds (ISGC) 2016 13-18 March 2016, Academia Sinica,
  Taipei, Taiwan}, Oct. 2016.
\newblock URL \url{https://doi.org/10.5281/zenodo.163537}.

\bibitem[{van Diepen} and {Dijkema}(2018)]{dppp}
G.~{van Diepen} and T.~J. {Dijkema}.
\newblock {DPPP: Default Pre-Processing Pipeline}.
\newblock Astrophysics Source Code Library, Apr. 2018.

\bibitem[Van~Haarlem et~al.(2013)Van~Haarlem, Wise, Gunst, Heald, McKean,
  Hessels, De~Bruyn, Nijboer, Swinbank, Fallows, et~al.]{LOFAR}
M.~Van~Haarlem, M.~Wise, A.~Gunst, G.~Heald, J.~McKean, J.~Hessels,
  A.~De~Bruyn, R.~Nijboer, J.~Swinbank, R.~Fallows, et~al.
\newblock {L}{O}{F}{A}{R}: The low-frequency array.
\newblock \emph{Astronomy \& astrophysics}, 556:\penalty0 A2, 2013.

\bibitem[Van~Weeren et~al.(2016)Van~Weeren, Williams, Hardcastle, Shimwell,
  Rafferty, Sabater, Heald, Sridhar, Dijkema, Brunetti,
  et~al.]{lofar_prefactor}
R.~Van~Weeren, W.~Williams, M.~Hardcastle, T.~Shimwell, D.~Rafferty,
  J.~Sabater, G.~Heald, S.~Sridhar, T.~Dijkema, G.~Brunetti, et~al.
\newblock L{O}{F}{A}{R} facet calibration.
\newblock \emph{The Astrophysical Journal Supplement Series}, 223\penalty0
  (1):\penalty0 2, 2016.

\bibitem[Virtanen et~al.(2018)Virtanen, Norden, Gunst, and
  Vermeulen]{staiton_data_cookbook}
I.~Virtanen, M.~Norden, A.~Gunst, and R.~Vermeulen.
\newblock Station data cookbook v1.2, 2018.
\newblock URL
  \url{http://lofar.ie/wp-content/uploads/2018/03/station_data_cookbook_v1.2.pdf}.

\bibitem[Williams et~al.(2016)Williams, Van~Weeren, R{\"o}ttgering, Best,
  Dijkema, de~Gasperin, Hardcastle, Heald, Prandoni, Sabater,
  et~al.]{Wendy_bootes}
W.~Williams, R.~Van~Weeren, H.~R{\"o}ttgering, P.~Best, T.~Dijkema,
  F.~de~Gasperin, M.~Hardcastle, G.~Heald, I.~Prandoni, J.~Sabater, et~al.
\newblock L{O}{F}{A}{R} 150-{M}{H}z observations of the {B}o{\"o}tes field:
  catalogue and source counts.
\newblock \emph{Monthly Notices of the Royal Astronomical Society},
  460\penalty0 (3):\penalty0 2385--2412, 2016.

\bibitem[Witt et~al.(2018)Witt, Bux, Gusew, and Leser]{Witt2018PredictivePM}
C.~Witt, M.~Bux, W.~Gusew, and U.~Leser.
\newblock Predictive performance modeling for distributed computing using
  black-box monitoring and machine learning.
\newblock \emph{CoRR}, abs/1805.11877, 2018.

\bibitem[Xu et~al.(1996)Xu, Zhang, and Sun]{semi_analytical_model}
Z.~Xu, X.~Zhang, and L.~Sun.
\newblock Semi-empirical multiprocessor performance predictions.
\newblock \emph{Journal of Parallel and Distributed Computing}, 39\penalty0
  (1):\penalty0 14 -- 28, 1996.
\newblock ISSN 0743-7315.
\newblock \doi{https://doi.org/10.1006/jpdc.1996.0151}.
\newblock URL
  \url{http://www.sciencedirect.com/science/article/pii/S0743731596901513}.

\bibitem[Yang et~al.(2005)Yang, Ma, and Mueller]{cross_platform_black_box}
L.~T. Yang, X.~Ma, and F.~Mueller.
\newblock Cross-platform performance prediction of parallel applications using
  partial execution.
\newblock In \emph{Proceedings of the 2005 ACM/IEEE conference on
  Supercomputing}, page~40. IEEE Computer Society, 2005.

\end{thebibliography}

\end{document}